\documentclass[10pt, draftclsnofoot, onecolumn]{IEEEtran}
\usepackage{amssymb}
\usepackage{cite}
\usepackage{graphicx}
\usepackage{array}
\usepackage{multirow}
\usepackage{amsmath}
\usepackage{stfloats}
\usepackage{graphicx}
\usepackage{subfigure}
\usepackage{tabularx}
\usepackage{epsfig,epsf,balance,cite}
\usepackage[caption=false,font=normalsize,labelfont=sf,textfont=sf]{subfig}
\usepackage{setspace}
\usepackage{bm}
\usepackage{textcomp}
\usepackage{algorithmic}
\usepackage{algorithm}
\usepackage{subfigure}
\usepackage{caption}
\usepackage{graphicx}
\usepackage{setspace}
\usepackage{url}
\usepackage{verbatim}
\usepackage{graphicx}
\usepackage{cite}
\usepackage{amssymb}
\usepackage{color}
\usepackage{xpatch}
\definecolor{shadecolor}{RGB}{0,0,255}
\definecolor{blue}{RGB}{0,0,255}

\newtheorem{theorem}{Theorem}

\newtheorem{lemma}{Lemma}

\makeatletter

\ExplSyntaxOn
\cs_new:Npn \bibColoredItems #1#2
{
	\clist_map_inline:nn {#2} { \cs_new:cpn {bib@colored@##1} {#1} } 
}
\ExplSyntaxOff

\newcommand\bib@setcolor[1]{%
	\ifcsname bib@colored@#1\endcsname
	\expanded{\noexpand\color{\csname bib@colored@#1\endcsname}}%
	\else
	\normalcolor
	\fi
}

\xpatchcmd\@bibitem
{\item}
{\bib@setcolor{#1}\item}
{}{\fail}

\xpatchcmd\@lbibitem
{\item}
{\bib@setcolor{#2}\item}
{}{\fail}
\makeatother

\begin{document}

\title{Resource Allocation for Cell-free Massive MIMO-enabled URLLC Downlink Systems}

\author{Qihao Peng, Hong Ren, Cunhua Pan, Nan Liu, and Maged Elkashlan
\thanks{Copyright (c) 2015 IEEE. Personal use of this material is permitted. However, permission to use this material for any other purposes must be obtained from the IEEE by sending a request to pubs-permissions@ieee.org.}
\thanks{Manuscript received May 31, 2022; revised October 11, 2022; accepted January 30, 2023. The work of Q. Peng was supported by the China Scholarship Council. This work was supported in part by the National Natural Science Foundation of China (Grant No. 62101128), National Key Research and Development Project (Grant No. 2019YFE0123600), National Natural Science Foundation of China (Grant No. 62201137), and Basic Research Project of Jiangsu Provincial Department of Science and Technology (Grant No. BK20210205).}
\thanks{Q. Peng and M. Elkashlan are with the School of Electronic Engineering and Computer Science at Queen Mary University of London, U.K. (e-mail: \{q.peng, maged.elkashlan\}@qmul.ac.uk). H. Ren, C. Pan, and N. Liu are with National Mobile Communications Research Laboratory, Southeast University, Nanjing, China. (e-mail:\{hren, cpan, nanliu\}@seu.edu.cn). ({\emph {Corresponding author: Hong Ren, Cunhua Pan}.})}
}

\maketitle

\begin{abstract}
Ultra-reliable and low-latency communication (URLLC) is a pivotal technique for enabling the wireless control over industrial Internet-of-Things (IIoT) devices. By deploying distributed access points (APs), cell-free massive multiple-input and multiple-output (CF mMIMO) has great potential to provide URLLC services for IIoT devices. In this paper, we investigate CF mMIMO-enabled URLLC in a smart factory. Lower bounds (LBs) of downlink ergodic data rate under finite channel blocklength (FCBL) with imperfect channel state information (CSI) are derived for maximum-ratio transmission (MRT), full-pilot zero-forcing (FZF), and local zero-forcing (LZF) precoding schemes. Meanwhile, the weighted sum rate is maximized by jointly optimizing the pilot power and transmission power based on the derived LBs. Specifically, we first provide the globally optimal solution of the pilot power, and then introduce some approximations to transform the original problems into a series of subproblems, which can be expressed in a geometric programming (GP) form that can be readily solved. Finally, an iterative algorithm is proposed to optimize the power allocation based on various precoding schemes. Simulation results demonstrate that the proposed algorithm is superior to the existing algorithms, and that \textcolor{black}{the quality of URLLC services will benefit by deploying more APs, except for the FZF precoding scheme.} 

\end{abstract}

\begin{IEEEkeywords}
Cell-free massive MIMO, URLLC, Industrial Internet-of-Things (IIoT).
\end{IEEEkeywords}

\section{Introduction}

Ultra-Reliable Low-Latency Communication (URLLC) is one of the crucial techniques in the next generation  industrial systems, which can support the mission-critical communication for industrial Internet-of-Things (IIOT) devices such as autonomous vehicles and robots \cite{2014Industrial,2019URLLC}. For industrial applications, the control command data packet size is generally small with the stringent requirements of low latency ($1$ $\rm ms$) and \textcolor{black}{low block error rate below $10^{-6}$} \cite{Hong-twc}. \textcolor{black}{Since the blocklength no longer tends to be infinite, the impact of the decoding error probability (DEP) should be considered. To investigate the coding rate in the short packet regime}, the authors of \cite{ref3} derived the approximated achievable data rate under finite channel blocklength (FCBL), which was expressed in a complex function of the channel blocklength and DEP \cite{ref1}. However, the achievable data rate expression is neither convex nor concave with respect to channel blocklength and signal-to-noise ratio (SNR) \cite{2017nonconvex}, which is challenging for resource allocation.


Recently, there are some contributions on resource allocation based on short packet transmission \cite{2019pan,2020Min,2021NOMA}. By deploying an unmanned aerial vehicle (UAV) as a relay, the short packet can be delivered to an obstructed device by optimizing UAV's location and channel blocklength \cite{2019pan}. The joint optimization on power allocation and blocklength was studied in \cite{2020Min}. The overall DEP was minimized by optimizing the power allocation in non-orthogonal multiple access (NOMA) systems \cite{2021NOMA}. \textcolor{black}{However, all the above studies \cite{2019pan,2020Min,2021NOMA} only considered a simple scenario with point-to-point link, while a smart industry needs to provide URLLC services for a large number of devices \cite{2018IOT}.} To support multiple devices, the orthogonal frequency division multiple access (OFDMA) technique was adopted in \cite{ghanem2019resource}, and the authors therein aimed to minimize the total bandwidth by optimizing the subchannel allocation. However, the frequency resource in IIoT applications is limited \cite{2018ind} and the OFDMA technique is not  effective for supporting an excessive number of devices.

Owing to a large number of available spatial degrees of freedom, massive multiple-input and multiple-output (mMIMO) can simultaneously support multiple devices by using the same time-frequency resources \cite{ref11,2014MIMO}, and thus the mMIMO-enabled URLLC has attracted extensive research attention \cite{ref13b1,ref13a,ref13b,ref13d,ref13}. The pilot length was optimized to minimize the DEP in \cite{ref13b1}, and the authors also analyzed the relationship between the latency and the DEP in mMIMO systems. Then, Zeng {\emph {et al}.} extended the results in \cite{ref13b1} to mMIMO systems with shadow fading, demonstrating that mMIMO can provide URLLC services for multiple devices even suffering from severe shadow fading \cite{ref13a}. The optimal secure performance was obtained by optimizing the channel blocklength and transmission bits per packet in \cite{ref13b}. The pilot power and payload power was jointly optimized to maximize the weighted sum rate of multiple devices in a single cell \cite{ref13d}. The authors of \cite{ref13} considered a more general scenario of multiple cells with imperfect channel state information (CSI) and pilot contamination, and showed that the pilot contamination had a significant impact on the reliability of URLLC services. Although it has been shown that mMIMO can provide URLLC services for multiple devices, \textcolor{black}{it may be unable to provide guaranteed URLLC services to all devices in the cell due to blockage issue and severe inter-cell interference.} Therefore, a novel network architecture should be developed to support URLLC services.

By geographically deploying distributed APs, cell-free mMIMO (CF mMIMO) can provide uniform services for all devices \cite{2018cell,ref14,2019ce}. The performance improvements of CF MIMO systems over the centralized mMIMO systems have been shown when using maximum ratio transmission (MRT) precoding scheme \cite{2017small} and zero forcing (ZF) precoding scheme \cite{2020liu}, respectively. Considering that the previous precoding schemes may no longer be applicable for CF mMIMO, the authors of \cite{ref18} proposed four precoding schemes, namely, full-pilot zero-forcing (FZF), local partial zero-forcing precoding, local protective partial zero-forcing, and local regularized zero-forcing. The energy efficiency of CF mMIMO was analyzed in \cite{2021EE}. The aforementioned works in \cite{2018cell,ref14,2019ce,2017small,2020liu,ref18} assumed that the APs can acquire the CSI of all devices, which is theoretically possible but impractical. To address this issues, a user-centric approach was proposed to reduce the implementation complexity \cite{ref18b}. The coverage probability with various densities of APs was analyzed in \cite{ref18c}. To tackle the blockage issue, the performance of the reconfigurable intelligent surface-aided CF mMIMO system was analyzed in \cite{2021CF_RIS}. However, all works were based on the assumption of infinite channel blocklength, which is not suitable for short packet transmission.

Due to the appealing advantages of CF mMIMO, it has great potential to provide URLLC services for multiple devices simultaneously in a large coverage area. Essentially, there was a significant improvement in terms of the network's availability over the centralized mMIMO \cite{CF21Per}. The power allocation based on FCBL for maximizing the minimal data rate and maximizing the energy efficiency was considered in \cite{ref20}, where each AP was equipped with a single antenna. \textcolor{black}{However, channel hardening can only be achieved by  deploying ultra-high density of single-antenna APs \cite{2018Channelharden}, which is theoretically possible but practically unrealistic due to the expensive hardware.} In this paper, \textcolor{black}{we investigate the deployment of multiple-antenna APs and optimal AP selection under the short packet regime}. Then, we aim to maximize the weighted sum rate based on FCBL while considering the minimal requirements of DEP and data rate, by optimizing the pilot power and the transmission power. The main contributions of this paper are summarized as follows.
\begin{enumerate}
  \item By using the user-centric approach, we derive the lower bounds (LBs) of the achievable downlink data rate with imperfect CSI for the MRT, FZF, and local zero-forcing (LZF) precoding schemes when using FCBL.
  \item The weighted sum rate is maximized by jointly optimizing the pilot power and the transmission power while considering the minimal requirements of DEP and data rate. To solve this NP-hard problem, \textcolor{black}{we first transform the DEP and data rate requirements into the required SINR, and then reducing the number of variables by proving that the globally optimal solution of pilot power can be derived in closed form. Furthermore}, by introducing the approximations, the problem can be simplified into a series of subproblems, which can be transformed into a geometric programming (GP) problem by using \textcolor{black}{log-function method} and successive convex approximation (SCA) \textcolor{black}{\cite{mehanna2014feasible,van2018joint}}. Finally, an iterative algorithm is proposed to solve this problem for three linear precoding schemes.
  \item Simulation results demonstrate the rapid convergence speed of our proposed algorithms, and also validate the effectiveness of our method over the existing algorithm. Besides, by accessing various APs, the optimal AP selection strategy based on short packet transmission is provided. More importantly, the CF mMIMO system has a remarkable performance improvement over the centralized mMIMO system.
\end{enumerate}

The remainder of this paper is organized as follows. In Section II, the system model is provided, and then the LB date rate expression under FCBL based on statistical CSI is derived for the MRT, FZF, and LZF precoding schemes, respectively. In Section III, the power allocation is optimized to maximize the ergodic sum data rate. Then, simulation results are presented in Section IV. Finally, the conclusions are drawn in Section V.

\textcolor{black}{\emph{Notation}: The superscripts $(\cdot)^{*}, (\cdot)^{T}, (\cdot)^{H}$ stand for the conjugate, transpose, and conjugate-transpose, respectively. The Euclidean norm and the expectation operator are denoted by $||\cdot||$ and $\mathbb{E}\left\lbrace \cdot \right\rbrace $, respectively. $z \sim {\mathcal{CN}} ({0,1})$ denotes a circularly symmetric complex Gaussian random variable (RV) $z$ with zero mean and unit variance, and ${\bf{z}} \sim {\mathcal{CN}} ({{\bf{0}},{\bf{I}}_N})$ means an $N$-dimensional complex vector, each element of which is independent and follows the distribution of $\mathcal{CN} \left( {{{0}},{{1}}} \right)$. Finally, ${\bf A} \in {\mathbb C}^{M \times N}$ means that $\bf A$ is a complex matrix with $M$ rows and $N$ columns. } 

\section{System Model and Spectral Efficiency}
\subsection{System Model}
\begin{figure}[t]
\centering
\includegraphics[width=3.2in]{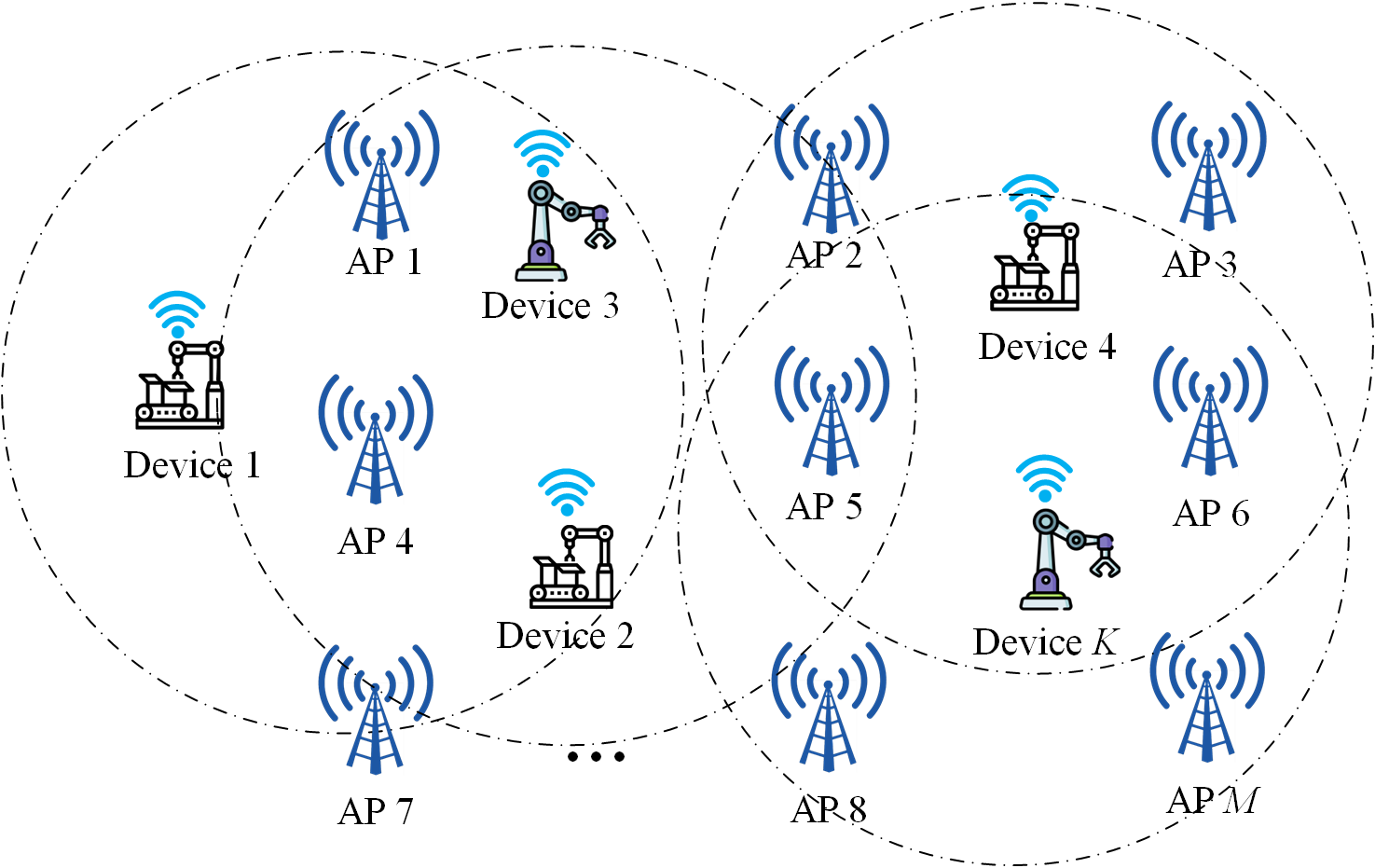}
\caption{Smart factory scenario where CF mMIMO serves multiple devices.}
\label{fig:system model}
\end{figure}

We consider a CF mMIMO-enabled smart factory where $M$ APs equipped with $N$ antennas jointly serve all $K$ single-antenna devices, as illustrated in Fig. \ref{fig:system model}. The channel vector ${\bf{g}}_{m,k} \in {\mathbb{C}}^{N \times 1} $ between the $m$th AP and the $k$th device is modeled as
\begin{equation}
\setlength\abovedisplayskip{5pt}
\setlength\belowdisplayskip{5pt}
\label{cofficient_g}
{{\bf{g}}_{m,k}} = \sqrt {{\beta _{m,k}}} {{\bf{h}}_{m,k}},
\end{equation}
where $\beta _{m,k}$ is the large-scale fading and \textcolor{black}{${\bf{h}}_{m,k} \sim \mathcal{CN} \left( {{\bf{0}},{\bf{I}}_N} \right)$ denotes a normal distribution with zero mean and variance of ${\bf{I}}_N$.}

\subsection{Uplink Training}
It is assumed that each AP needs to estimate the CSI from all the devices based on time division duplex (TDD) protocol within the limited channel blocklength $L = B \times T_B$, where $B$ is the bandwidth and $T_B$ is the transmission duration. In order to distinguish the channels from different devices, $K$ devices are allocated with orthogonal pilot sequences. Then, the $m$th AP estimates the channel matrix based on the received pilot signal \textcolor{black}{${{\bf{Y}}^p_{m}} \in {{\mathbb{C}}^{N \times K}}$}, which is given by
\begin{equation}
\label{received_pilot}
\setlength\abovedisplayskip{5pt}
\setlength\belowdisplayskip{5pt}
{{\bf{Y}}^p_{m}} = \sum\limits_{k = 1}^K {{{\bf{g}}_{m,k}}\sqrt {{K}p_k^p} {\bf{ q }}_k^H}  + {{\bf{N}}^p_{m}},
\end{equation}
where $p_k^p$ is the pilot power of the $k$th device, \textcolor{black}{${\bf{ q }}_k \in {{\mathbb{C}}^{K \times 1}}$ is the $k$th device's pilot sequence}, and \textcolor{black}{${{\bf{N}}_{m}^p} \in  {{\mathbb{C}}^{N \times K}}$} is the additive Gaussian noise matrix at the $m$th AP, each element of which is independent and follows the distribution of $\mathcal{CN} \left( {{{0}},{{1}}} \right)$. By multiplying (\ref{received_pilot}) with orthogonal pilot ${\bf{ q }}_k$, we have
\begin{equation}
\label{received_channel}
\setlength\abovedisplayskip{5pt}
\setlength\belowdisplayskip{5pt}
{{\bf{\hat y}}_{m,k}^p} = \frac{1}{{\sqrt {Kp_k^p} }}{{\bf{Y}}^p_{m}}{{\bf{q}}_k} = {{\bf{g}}_{m,k}} + {\bf{n}}_{m,k}^p,
\end{equation}
where ${\bf{n}}_{m,k}^p = \frac{1}{{\sqrt {Kp_k^p} }}{{\bf{N}}^p_{m}}{{\bf{q}}_k}$. Based on (\ref{received_channel}), the estimated channel vector  ${\bf{\hat g}}_{m,k}$ by \textcolor{black}{using minimum mean-square error (MMSE)} is
\begin{equation}
\label{estimate_gmk}
\setlength\abovedisplayskip{5pt}
\setlength\belowdisplayskip{5pt}
{{{\bf{\hat g}}}_{m,k}} = \frac{{K{p_{k}^p}{\beta _{m,k}}}}{{K{p_{k}^p}{\beta _{m,k}} + 1}}{{{\bf{\hat y}}}^p_{m,k}},
\end{equation}
which follows the distribution of $\mathcal{CN} \left( {{{\bf {0}}},{{{\lambda_{m,k}}{{\bf{I}}_N}}}} \right)$ with $\lambda_{m,k}$ given by
\begin{equation}
\label{gama_mk}
\setlength\abovedisplayskip{5pt}
\setlength\belowdisplayskip{5pt}
{\lambda_{m,k}} = \frac{{Kp_k^p{{\left( {{\beta _{m,k}}} \right)}^2}}}{{Kp_k^p{\beta _{m,k}} + 1}}.
\end{equation}
Then, let us denote ${{{\bf{\tilde g}}}_{m,k}} = {{\bf{g}}_{m,k}} - {{{\bf{\hat g}}}_{m,k}}$ as the channel estimation error, which is independent of $ {{{\bf{\hat g}}}_{m,k}}$ and follows the distribution of $\mathcal{CN} \left( {{{\bf{0}}},{{ \left( {{\beta _{m,k}} - {\lambda _{m,k}}} \right){{\bf{I}}_N}}}} \right)$.

\subsection{Downlink Transmission}
For downlink transmission, to reduce the computational complexity, the user-centric approach is adopted, e.g., each device is served by a subset of APs or each AP serves a subset of devices. Denote ${\mathcal{M}}_k$ as the set of APs that serve the $k$th device and ${\mathcal{U}}_m$ as the set of devices that are served by the $m$th AP, respectively. The transmitted signal from the $m$th AP is denoted as
\begin{equation}
\label{transmitted_signal}
\setlength\abovedisplayskip{5pt}
\setlength\belowdisplayskip{5pt}
{{\bf{x}}_m} = \sum\limits_{k \in {{\cal U}_m}} {\sqrt {p_{m,k}^d} {{\bf{a}}_{m,k}^ *}{s_k}},
\end{equation}
where ${p_{m,k}^d}$ is the transmission power, ${{\bf{a}}_{m,k}}$ is the precoding vector, and $s_k$ is the data symbol to the $k$th device.

The received signal at the $k$th device is
\begin{align}
\setlength\abovedisplayskip{5pt}
\setlength\belowdisplayskip{5pt}
y_k^d &= \sum\limits_{m = 1}^M {\sum\limits_{k' \in {{\cal U}_m}} {\sqrt {p_{m,k'}^d} {\bf{g}}_{m,k}^T{\bf{a}}_{m,k'}^ * {s_{k'}}} }  + {n_k} \notag \\
& = \sum \limits_{k' = 1}^K {\sum\limits_{m \in {{\cal M}_{k'}}} {{{\left( {{{\bf{g}}_{m,k}}} \right)}^T}{\bf{a}}_{m,k'}^ * \sqrt {p_{m,k'}^d} {s_{k'}}} }  + {n_k}, \label{downlink_kth_signal}
\end{align}
where $n_k$ is the noise with the distribution of $\mathcal{CN} \left( {{{0}},{{1}}} \right)$. Besides, since there are no downlink pilots, we assume that the $k$th device treats the mean of the effect channel gain as the true channel for signal detection \cite{2018CSI}. Then, the received signal at the $k$th device can be rewritten as
\begin{align}
\setlength\abovedisplayskip{5pt}
\setlength\belowdisplayskip{5pt}
y_k^d & = \underbrace { {\mathbb E}\left\{ {\sum\limits_{m \in {{\cal M}_k}} {{{\left( {{{\bf{g}}_{m,k}}} \right)}^T}{\bf{a}}_{m,k}^ * \sqrt {p_{m,k}^d} } } \right\}}_{{\rm{D}}{{\rm{S}}_k}}{s_k}  \notag \\
& + \underbrace {\left\{ {\sum\limits_{m \in {{\cal M}_k}} {{{\left( {{{\bf{g}}_{m,k}}} \right)}^T}{\bf{a}}_{m,k}^ * \sqrt {p_{m,k}^d} }  - {{\rm{D}}{{\rm{S}}_k}}} \right\}}_{{\rm LS}_{k}}{s_k} , \label{downlink_kth_statistics_signal} \\
& + \sum\limits_{k' \ne k}^K {\underbrace {\sum\limits_{m \in {{\cal M}_{k'}}} {{{\left( {{{\bf{g}}_{m,k}}} \right)}^T}{{ {{\bf{a}}_{m,k'}^ * } } }\sqrt {p_{m,k'}^d} } }_{{\rm UI}_{{k,k'}}}{s_{k'}}}  + \underbrace {{n_k}}_{{{\rm N}_k}}, \notag
\end{align}
where ${\rm{DS}}_k$ is the desired signal, ${\rm {LS}}_k$ is the leaked signal, ${\rm {UI}}_{k,k'}$ represents the interference due to the $k'$th device, and ${\rm {N}}_k$ is the noise term. The SINR of the $k$th device is given by
\begin{equation}
\setlength\abovedisplayskip{5pt}
\setlength\belowdisplayskip{5pt}
\label{kth_SINR_downlink}
\gamma _k = \frac{{{{\left| {{\rm{DS}}_{k}} \right|}^2}}}{{{{\left| {{\rm{LS}}_{k}} \right|}^2} + \sum\nolimits_{k' \ne k}^K {{{\left| {{\rm{UI}}_{k,k'}} \right|}^2}}  + {{\left| {{{\rm{N}}_k}} \right|}^2}}}.
\end{equation}

For the precoding vector ${{\bf{a}}_{m,k}}$, we consider the following three linear precoding schemes \textcolor{black}{\cite{ref18,zhang2020prospective}}
\begin{equation}
\setlength\abovedisplayskip{5pt}
\setlength\belowdisplayskip{5pt}
\label{precoding_downlink}
{{\bf{a}}_{m,k}} = \left\{ {\begin{array}{*{20}{c}}
\frac{{{{{\bf{\hat G}}}_m}{{\bf{e}}_k}}}{{\sqrt { \mathbb{E} \left\{ {{{\left\| {{{{\bf{\hat G}}}_m}{{\bf{e}}_k}} \right\|}^2}} \right\}} }},&{{\rm{MRT}}}\\
\frac{{{{{\bf{\hat G}}}_m}{{\left( {{\bf{\hat G}}_m^H{{{\bf{\hat G}}}_m}} \right)}^{ - 1}}{{\bf{e}}_k}}}{{\sqrt {\mathbb{E} \left\{ {{{\left\| {{{{\bf{\hat G}}}_m}{{\left[ {{\bf{\hat G}}_m^H{{{\bf{\hat G}}}_m}} \right]}^{ - 1}}{{\bf{e}}_k}} \right\|}^2}} \right\}} }},&{{\rm{FZF}}}\\
\frac{{{{{\bf{\hat G}}}_m}{{\bf{E}}_{{\mathcal {U}_m}}}{{\left( {{\bf{E}}_{{\mathcal {U}_m}}^H{\bf{\hat G}}_m^H{{{\bf{\hat G}}}_m}{{\bf{E}}_{{\mathcal {U}_m}}}} \right)}^{ - 1}}{\bm{\xi} _{m,k}}}}{{\sqrt {\mathbb{E}\left\{ {{{\left\| {{{{\bf{\hat G}}}_m}{{\bf{E}}_{{\mathcal {U}_m}}}{{\left( {{\bf{E}}_{{\mathcal {U}_m}}^H{\bf{\hat G}}_m^H{{{\bf{\hat G}}}_m}{{\bf{E}}_{{\mathcal {U}_m}}}} \right)}^{ - 1}}{\bm{\xi} _{m,k}}} \right\|}^2}} \right\}} }},&{{\rm{LZF}}}
\end{array}} \right.
\end{equation}
where $\mathbb{E}\left \{ \cdot \right \}$ denotes the expectation operator, ${\bf{\hat G}}_m = \left[ {{{{\bf{\hat g}}}_{m,1}},{{{\bf{\hat g}}}_{m,2}}, \cdot  \cdot  \cdot ,{{{\bf{\hat g}}}_{m,K}}} \right]$ is the estimated channel matrix between all the devices and the $m$th AP, and ${\bf{e}}_{k}$ represents the $k$th column of unit matrix ${\bf {I}}_K$. For the LZF precoding scheme, ${\bf{\hat G}}_m {{\bf{E}}_{{{\cal U}_m}}} = [{\bf{\hat g}}_{m,d_1},{\bf{\hat g}}_{m,d_2},\cdot \cdot \cdot, {\bf{\hat g}}_{m,d_{|{\cal{U}}_m|}}] \in {\mathbb C}^{N \times |{\cal U}_m|}$ is a matrix collecting the channels of serving \textcolor{black}{devices} in ${\cal{U}}_m$, where ${\cal{U}}_m =  \left\{ d_1, d_2,\cdot \cdot \cdot,{d_{|{\cal U}_m|}}\right\} $ is the set of devices served by the $m$th AP and ${{\bf{E}}_{{{\cal U}_m}}}$ is $\left[ {\bf{e}}_{d_1},{\bf{e}}_{d_2},\cdot \cdot \cdot,{\bf{e}}_{d_{|{\cal U}_m|}} \right] \in {\mathbb C}^{K \times |{\cal U}_m|}$. For ease of exposition, let $ {\cal U}_{m}^{\text{index}}  = \left\lbrace 1,2,\cdot \cdot \cdot, |{{\cal U}_m}|\right\rbrace $ be the set comprised of the index of ${\cal U}_m$. Given user $k$, we can find an index $j\in  {\cal U}_{m}^{\text{index}} $ where $d_j =k$. Then, we have ${\bm{\xi}}_{m,k} = \left[\mathbf{I}_{|{\cal{U}}_m|}\right]_{({:,j})}$.


As can be seen from (\ref{precoding_downlink}), for the FZF precoding scheme, the $m$th AP needs to estimate all devices' channels, and thus it can suppress the interference of all devices by sacrificing spatial degrees of freedom. In contrast, the $m$th AP using the MRT and the LZF precoding methods only needs to know the serving devices' CSI, which reduces the implementation complexity. Besides, the system based on  the LZF precoder can only suppresses the interference causing by serving devices, which strikes a balance between the available spatial degrees of freedom and the interference suppression.

\subsection{Achievable Data Rate under Finite Blocklength}
Based on Shannon's coding theorem, the Shannon capacity is defined as the maximum coding rate that there exists an encoder/decoder pair that can enable the DEP to approach zero when the channel blocklength is infinity \cite{1948Shanon}. However, in short packet transmission, the DEP has a non-negligible impact on the data rate. In \cite{ref3}, the authors derived the approximate achievable data rate for the $k$th device under FCBL, which is given by
\begin{equation}
\label{urllc_rate}
\setlength\abovedisplayskip{5pt}
\setlength\belowdisplayskip{5pt}
{R_k} \approx \left( {1 - \eta } \right){\log _2}\left( {1 + \gamma_k} \right) - \sqrt {\frac{{\left( {1 - \eta } \right){V_k\left(\gamma_k\right)}}}{L}} \frac{{{Q^{ - 1}}\left( {{\varepsilon _k}} \right)}}{{\ln 2}},
\end{equation}
where $\eta = K / L$, ${\gamma}_k$ is the $k$th device's SINR, ${{\varepsilon _k}}$ is DEP,  $V_k$ is the channel dispersion with  ${V_k\left(\gamma_k\right)} = 1 - {\left( {1 + {\gamma_k}} \right)^{ - 2}}$, and ${Q^{ - 1}}\left( {{\varepsilon _k}} \right)$ is the inverse function of $Q\left( {{\varepsilon _k}} \right) = \frac{1}{{\sqrt {2\pi } }}\int_{{\varepsilon _k}}^\infty  {{{\rm{e}}^{{{ - {t^2}} \mathord{\left/
 {\vphantom {{ - {t^2}} 2}} \right.
 \kern-\nulldelimiterspace} 2}}}{\rm{d}}t}$ of the $k$th device.

The ergodic data rate of the $k$th device under FCBL is given by
\begin{equation}
\setlength\abovedisplayskip{5pt}
\setlength\belowdisplayskip{5pt}
\label{rw_rate}
\begin{split}
{\bar R_k} & \approx \mathbb{E} \left\{ \frac{{1 - \eta }}{{\ln 2}}\left[ {\ln \left( {1 \!+\! {\gamma _k}} \right) \!-\! \frac{{{Q^{ - 1}}\left( {{\varepsilon _k}} \right)}}{{\sqrt {L\left( {1 - \eta } \right)} }}\!\sqrt {\frac{{\frac{2}{{{\gamma _k}}} + 1}}{{{{\left( {\frac{1}{{{\gamma _k}}} + 1} \right)}^2}}}} } \right] \right\}, \\
&\triangleq \frac{{1 - \eta }}{{\ln 2}} \mathbb{E} \left\{ f_k \left( \frac{1}{\gamma_k} \right)\right\},
\end{split}
\end{equation}
where \textcolor{black}{$f_k(x) = \ln(1+\frac{1}{x}) - \frac{{{Q^{ - 1}}\left( {{\varepsilon _k}} \right)}}{{\sqrt {L\left( {1 - \eta } \right)} }} \sqrt{\frac{2x + 1}{(1+x)^2}}$ is a function based on the $k$th device's DEP requirements, and} the expectation is taken over the small-scale fading channel. As can be seen from (\ref{rw_rate}), the closed-form expression of the ergodic data rate is challenging to derive, and thus we cannot allocate the power based on the exact expression of (\ref{rw_rate}). To address this issue, we aim to derive the LB of the ergodic data rate which is more convenient for resource allocation.

Assuming that the data rate $R_k$ of any device is no smaller than $0$, we have the following inequality
\begin{equation}
\setlength\abovedisplayskip{5pt}
\setlength\belowdisplayskip{5pt}
\label{a_region}
\frac{{{Q^{ - 1}}\left( {{\varepsilon _k}} \right)}}{{\sqrt {L\left( {1 - \eta } \right)} }} \le \frac{{\left( {\frac{1}{{{\gamma _k}}} + 1} \right)\ln \left( {1 + {\gamma _k}} \right)}}{{\sqrt {\frac{2}{{{\gamma _k}}} + 1} }} \buildrel \Delta \over =  g\left( \frac{1}{{\gamma}_k} \right ),
\end{equation}
\textcolor{black}{where $g(x)$ is equal to $\frac{(1+x)\ln(1+\frac{1}{x})}{\sqrt{2x + 1}}$}.  
We can readily check that the first-order derivative of $g\left( x \right )$ is smaller than $0$, and thus $g\left( x \right )$ is a monotonically decreasing function. Besides, the feasible region of $f_{k} \left(x \right)$ is $0 \le x \le {g^{ - 1}}\left( {\frac{{{Q^{ - 1}}\left( {{\varepsilon _k}} \right)}}{{\sqrt {L\left( {1 - \eta } \right)} }}} \right)$. As a result, we have the following lemma.

\begin{lemma}
\label{x_region}
Function $f_{k} \left(x \right)$ is a decreasing and convex function when $ 0 < x \le {g^{ - 1}}\left( {\frac{{{Q^{ - 1}}\left( {{\varepsilon _k}} \right)}}{{\sqrt {L\left( {1 - \eta } \right)} }}} \right) $. 

{\emph{Proof}}: Please refer to Appendix B in \cite{ref20a}. $\hfill\blacksquare$

\end{lemma}

By using Jensen's inequality and Lemma \ref{x_region}, we have
\begin{equation}
\setlength\abovedisplayskip{5pt}
\setlength\belowdisplayskip{5pt}
\label{rate}
{\bar R_k} \ge {\hat R}_k \triangleq \frac{{1 - \eta }}{{\ln 2}}{f_k}\left( {{1 \mathord{\left/
 {\vphantom {1 {{{\hat \gamma }_k}}}} \right.
 \kern-\nulldelimiterspace} {{{\hat \gamma }_k}}}} \right),
\end{equation}
where ${\hat R}_k$ is the LB data rate of the $k$th device, and ${{\hat \gamma }_k}$ is ${{\hat \gamma }_k} = \frac{1}{\mathbb{E} \left( {{1 \mathord{\left/
 {\vphantom {1 {{{\hat \gamma }_k}}}} \right.
 \kern-\nulldelimiterspace} {{{ \gamma }_k}}}} \right)}$.

To obtain the closed-form expression of ${\hat R}_k$, the $k$th device's SINRs based on the MRT, FZF, and LZF precoding schemes should be derived. Specifically, we have the following results.

\begin{theorem}
\label{MRC_SINR_T}
The ergodic achievable data rate for the $k$th device using the MRT precoding scheme under FCBL can be lower bounded by
\begin{equation}
\setlength\abovedisplayskip{5pt}
\setlength\belowdisplayskip{5pt}
\label{MRC_LB_rate}
 {\hat R}_k^{\rm MRT} \triangleq  \frac{{1 - \eta }}{{\ln 2}}f_k \left( \frac{1}{{\hat \gamma }_k^{\rm MRT}} \right),
\end{equation}
where ${\hat \gamma }_k^{\rm MRT}$ is denoted as
\begin{equation}
\setlength\abovedisplayskip{5pt}
\setlength\belowdisplayskip{5pt}
\label{MRC_SINR_LB}
\hat \gamma _k^{{\rm{MRT}}} = \frac{{{{\left( {\sum\limits_{m \in {{\cal M}_k}} {\sqrt {N p_{m,k}^d{\lambda _{m,k}}} } } \right)}^2}}}{{\sum\limits_{k' = 1}^K {\sum\limits_{m \in {{\cal M}_{k'}}} {p_{m,k'}^d{\beta _{m,k}}} }  + 1}}.
\end{equation}

{\emph{Proof}}: Please refer to Appendix \ref{MRC_SINR_P}. $\hfill\blacksquare$

\end{theorem}

\begin{theorem}
\label{FZF_SINR_T}
Using the FZF precoding scheme, the $k$th device's ergodic data rate is lower bounded by
\begin{equation}
\setlength\abovedisplayskip{5pt}
\setlength\belowdisplayskip{5pt}
\label{FZF_LB_rate}
 {\hat R}_k^{\rm FZF} \triangleq  \frac{{1 - \eta }}{{\ln 2}} f_k \left( \frac{1}{{\hat \gamma }_k^{\rm FZF}} \right),
\end{equation}
where ${\hat \gamma }_k^{\rm FZF}$ is denoted as
\begin{equation}
\setlength\abovedisplayskip{5pt}
\setlength\belowdisplayskip{5pt}
\label{SINR_ZF}
\hat \gamma _k^{{\rm{FZF}}} = \frac{{{{\left( { \sum\limits_{m \in {{\cal M}_k}} {\sqrt {{(N - K)} p_{m,k}^d{\lambda _{m,k}}} } } \right)}^2}}}{{\sum\limits_{k' = 1}^K {\sum\limits_{m \in {{\cal M}_{k'}}} {p_{m,k'}^d\left( {{\beta _{m,k}} - {\lambda _{m,k}}} \right)} }  + 1}},
\end{equation}
where the number of antennas $N$ should be larger than the number of devices $K$.

{\emph{Proof}}: Please refer to Appendix \ref{FZF_SINR_P}. $\hfill\blacksquare$

\end{theorem}

\begin{theorem}
\label{LZF_SINR_T}
The $k$th device's ergodic data rate based on the LZF precoding scheme is lower bounded by
\begin{equation}
\setlength\abovedisplayskip{5pt}
\setlength\belowdisplayskip{5pt}
\label{LZF_LB_rate}
 {\hat R}_k^{\rm LZF} \triangleq  \frac{{1 - \eta }}{{\ln 2}} f_k \left( \frac{1}{{\hat \gamma }_k^{\rm LZF}} \right),
\end{equation}
where ${\hat \gamma }_k^{\rm LZF}$ is given by
\begin{equation}
\setlength\abovedisplayskip{5pt}
\setlength\belowdisplayskip{5pt}
\label{SINR_LZF}
\hat \gamma _k^{{\rm{LZF}}} = \frac{{{{\left( {\sum\limits_{m \in {{\cal M}_k}} {\sqrt {\left( {N - {\tau _m}} \right)p_{m,k}^d{\lambda _{m,k}}} } } \right)}^2}}}{{\sum\limits_{k' = 1}^K {\left[ {\sum\limits_{m \in \left\{ {{{\cal M}_{k'}} \cap {{\cal M}_k}} \right\}} {p_{m,k'}^d\left( {{\beta _{m,k}} - {\lambda _{m,k}}} \right)}  + \sum\limits_{m \in \left\{ {{{\cal M}_{k'}}\backslash \left\{ {{{\cal M}_k} \cap {{\cal M}_{k'}}} \right\}} \right\}} {p_{m,k'}^d{\beta _{m,k}}} } \right]}  + 1}}.
\end{equation}

In (\ref{SINR_LZF}), ${\tau _m}$ means the number of devices served by the $m$th AP and its value is given by ${\tau _m} = |{\mathcal{U}}_m|$. Here, the number of antennas $N$ should be larger than ${\tau _m}$.

{\emph{Proof}}: Please refer to Appendix \ref{LZF_SINR_P}. $\hfill\blacksquare$
\end{theorem}

From the expressions of SINRs in (\ref{SINR_ZF}) and (\ref{SINR_LZF}), the FZF precoding scheme is a special case of the LZF precoder, i.e., the device is served by all APs. Besides, we also note that $\tau_m$ is always no larger than the number of devices $K$, as $\mathcal{U}_m$ is a subset of devices. Therefore, by choosing \textcolor{black}{the} optimal set of APs, it is reasonable for the system to adopt the LZF precoding scheme to support more devices than that based on the FZF precoder.

\section{Power Allocation}
In this section, we aim to optimize the power allocation to maximize the weighted sum rate.

\subsection{Problem Formulation}
We assume that all the devices have the same bandwidth $B$, and we aim to maximize the weighted sum rata with limited energy constraints and \textcolor{black}{the} minimal data rate requirement. Mathematically, the optimization problem can be formulated as
\begin{subequations}
\setlength\abovedisplayskip{5pt}
\setlength\belowdisplayskip{5pt}
\label{MRC_optimization}
\begin{align}
\mathop {\max }\limits_{\left\{ {p_k^p} \right\},\left\{ {p_{m,k}^d} \right\}} & \sum\limits_{k = 1}^K {{w_k}{{\hat R}_k} } \label{MRC_optimization_a}\\
{\rm{s}}{\rm{.t}}{\rm{.}}\;\;\;\; & {{\hat R}_k } \ge R_k^{{\rm{req}}},\forall k,  \label{MRC_optimization_b}\\
& {p_k^p \le P_k^{\max ,p}},\forall k  \label{MRC_optimization_c}\\
& {\sum\limits_{k \in {{\cal U}_m}}p_{m,k}^d \le {P_m^d},\forall m}, \label{MRC_optimization_d}
\end{align}
\end{subequations}
where ${{\hat R}_k}$ denotes the LB data rate based on the abovementioned three precoding schemes, $R_k^{{\rm{req}}}$ is the $k$th device's data rate requirement, $w_k$ is the weight of the $k$th device, $P_k^{\max ,p}$ is the maximal power of the $k$th device, $P_m^d$ is the $m$th AP's maximal transmission power. Specifically, constraint (\ref{MRC_optimization_b}) means the $k$th device's minimal data rate requirements, constraint (\ref{MRC_optimization_c}) and constraint (\ref{MRC_optimization_d}) mean that the uplink training power of each device and the total transmission power of each AP are limited.

\textcolor{black}{For the power allocation based on infinite blocklength in \cite{zhang2018performance,luo2022downlink}, the problem can be converted into a convex problem by introducing slack variables, which can be readily solved by the bisection search algorithm. However, maximizing the weighted sum rate is an NP-hard problem, which cannot be readily solved}. Besides, it is more challenging to solve the weighted sum rate problem under imperfect CSI and FCBL. Therefore, we first simplify the problem, and then propose an efficient algorithm for solving the problem with polynomial-time complexity.

Using Lemma \ref{x_region}, the minimal data rate requirement in (\ref{MRC_optimization_b}) can be transformed into the $k$th device's requirement of SINR, denoted as
\begin{equation}
\setlength\abovedisplayskip{5pt}
\setlength\belowdisplayskip{5pt}
\label{transformation}
\hat \gamma _k \ge \frac{1}{{f_k^{ - 1}\left( {\frac{{R_k^{{\rm{req}}}\ln 2}}{{1 - \eta }}} \right)}},
\end{equation}
where $\hat \gamma _k$ represents the $k$th device's SINR using the abovementioned precoding schemes. Besides, we find the globally optimal solution for pilot power based on the following lemma.

\begin{lemma}
\label{maximal_pp}
$f_{k}(\frac{1}{\hat \gamma_k})$ is a monotonically increasing function of pilot power $p_k^p$ when $ 0 < \frac{1}{\hat \gamma_k} \le {g^{ - 1}}\left( {\frac{{{Q^{ - 1}}\left( {{\varepsilon _k}} \right)}}{{\sqrt {L\left( {1 - \eta } \right)} }}} \right) $. 

{\emph{Proof}}: Please refer to Appendix \ref{proof_pp}. $\hfill\blacksquare$

\end{lemma}

By using (\ref{transformation}) and substituting $p_k^p = P_k^{{\rm max},p}$ into the SINR's expression, Problem (\ref{MRC_optimization}) can be simplified as
\begin{subequations}
\setlength\abovedisplayskip{5pt}
\setlength\belowdisplayskip{5pt}
\label{MRC_optimization_sim}
\begin{align}
\mathop {\max }\limits_{\left\{ {p_{m,k}^d} \right\}} & \sum\limits_{k = 1}^K {{w_k}{{\hat R}_k}} \label{MRC_optimization_sim_a}\\
{\rm{s}}{\rm{.t}}{\rm{.}}\;\;\;\; & \hat \gamma _k \ge \frac{1}{{f_k^{ - 1}\left( {\frac{{R_k^{{\rm{req}}}\ln 2}}{{1 - \eta }}} \right)}},\forall k,  \label{MRC_optimization_sim_b}\\
& {\sum\limits_{k \in {{\cal U}_m}}p_{m,k}^d \le {P_m},\forall m}. \label{MRC_optimization_sim_c}
\end{align}
\end{subequations}

Then, by introducing slack variables ${\chi _k}$, Problem (\ref{MRC_optimization}) can be equivalently transformed into the following optimization problem
\begin{subequations}
\setlength\abovedisplayskip{5pt}
\setlength\belowdisplayskip{5pt}
\label{MRC_optimization_sim_trans}
\begin{align}
\mathop {\max }\limits_{\left\{ {p_{m,k}^d} \right\},\left\{ {\chi_k} \right\}} & {\sum\limits_{k = 1}^K {{w_k}\frac{{\left( {1 - \eta } \right)}}{{\ln 2}}\left[ {\ln \left( {1 + {\chi _k}} \right) - {\alpha _k} G\left( {{\chi _k}} \right)} \right]} } \label{MRC_optimization_sim_trans_a} \\
\text{s.t.} \;\;\;\;\; & \hat \gamma _k \ge {\chi _k},\forall k,  \label{MRC_optimization_sim_trans_b}\\
&{\chi _k} \ge \frac{1}{{f_k^{ - 1}\left( {\frac{{R_k^{{\rm{req}}}\ln 2}}{{1 - \eta }}} \right)}},\forall k, \label{MRC_optimization_sim_trans_c} \\
&{\rm{ \left(\ref{MRC_optimization_sim_c}\right)}}, \label{MRC_optimization_sim_trans_d}
\end{align}
\end{subequations}
where $G\left( {{\chi _k}} \right)$ is defined as $G\left( {{\chi _k}} \right) \triangleq \sqrt {\frac{{\frac{2}{{{\chi _k}}} + 1}}{{{{\left( {\frac{1}{{{\chi _k}}} + 1} \right)}^2}}}}$, and $\alpha_k$ is $\alpha_k = {\frac{{{Q^{ - 1}}\left( {{\varepsilon _k}} \right)}}{{\sqrt {L\left( {1 - \eta } \right)} }}}$.

To further simplify the objective function in (\ref{MRC_optimization_sim_trans_a}), the following lemmas are introduced.

\begin{lemma}
\label{lnx}
For any given ${\hat {x}} \ge 0$, function $\ln\left(1 + x\right)$ can be lower bounded by
\begin{equation}
\setlength\abovedisplayskip{5pt}
\setlength\belowdisplayskip{5pt}
\label{lemma1}
\ln \left( {1 + x} \right) \ge \rho \ln x + \delta ,
\end{equation}
where $\rho$ and $\delta$ are expressed as
\begin{equation}
\setlength\abovedisplayskip{5pt}
\setlength\belowdisplayskip{5pt}
\label{rho}
\rho  = \frac{{\hat x}}{{1 + \hat x}},\delta  = \ln \left( {1 + \hat x} \right) - \frac{{\hat x}}{{1 + \hat x}}\ln \left( {\hat x} \right).
\end{equation}

{\emph{Proof}}: Please refer to Appendix \ref{proof_lnx}. $\hfill\blacksquare$

\end{lemma}

\begin{lemma}
\label{vx}
 For any given ${\hat {x}} \ge \frac{{\sqrt {17}  - 3}}{4}$, function $G\left(x \right)$ always satisfies the following inequality
\begin{equation}
\setlength\abovedisplayskip{5pt}
\setlength\belowdisplayskip{5pt}
\label{lemma2}
G\left( x \right) \le \hat \rho \ln \left( x \right) + \hat \delta ,
\end{equation}
where $\hat\rho$ and $\hat \delta $ are given by
\begin{equation}
\setlength\abovedisplayskip{5pt}
\setlength\belowdisplayskip{5pt}
\label{tilderho}
\hat \rho  = \frac{{\hat x}}{{\sqrt {{{\hat x}^2} + 2\hat x} }} - \frac{{\hat x\sqrt {{{\hat x}^2} + 2\hat x} }}{{{{\left( {1 + \hat x} \right)}^2}}},
\end{equation}

and
\begin{equation}
\setlength\abovedisplayskip{5pt}
\setlength\belowdisplayskip{5pt}
\label{tildedelta}
\hat \delta  = \sqrt {1 - \frac{1}{{{{\left( {1 + \hat x} \right)}^2}}}}  - \hat \rho \ln \left( {\hat x} \right).
\end{equation}

{\emph{Proof}}: Please refer to Appendix D in \cite{ref13d}. $\hfill\blacksquare$

\end{lemma}

By using Lemma \ref{lnx} and Lemma \ref{vx}, the weighted data rate can be approximated in an iterative manner, which is detailed as follows
\begin{align}
\setlength\abovedisplayskip{5pt}
\setlength\belowdisplayskip{5pt}
& {{w_k}\frac{{\left( {1 - \eta } \right)}}{{\ln 2}}\left[ {\ln \left( {1 + {\gamma _k}} \right) - {\alpha _k}G\left( {{\gamma _k}} \right)} \right]} \notag \\
\ge & {{w_k}\frac{{\left( {1 - \eta } \right)}}{{\ln 2}}\left[ {{\rho ^{\left( i \right)}_{k}}\ln \left( {{\gamma _k}} \right) + {\delta ^{\left( i \right)}_{k}} - {\alpha _k}{{\hat \rho }^{\left( i \right)}_{k}}\ln \left( {{\gamma _k}} \right) - {\alpha _k}{{\hat \delta }^{\left( i \right)}_{k}}} \right]}, \label{wk_appro}
\end{align}
where ${\rho ^{\left( i \right)}_{k}}$, ${\delta ^{\left( i \right)}_{k}}$, ${{\hat \rho }^{\left( i \right)}_{k}}$, and ${{\hat \delta }^{\left( i \right)}_{k}}$ are obtained based on ({\ref{rho}), ({\ref{tilderho}), and ({\ref{tildedelta}) by using $\hat x={\gamma}_k^{\left( i \right)}$ in the $i$th iteration. As a result, the weighted sum rate in (\ref{MRC_optimization_sim_trans_a}) can be lower bounded by
\begin{align}
\setlength\abovedisplayskip{5pt}
\setlength\belowdisplayskip{5pt}
& \sum\limits_{k = 1}^K {{w_k}\frac{{\left( {1 - \eta } \right)}}{{\ln 2}}\left[ {\ln \left( {1 + {\chi _k}} \right) - {\alpha _k}G\left( {{\chi _k}} \right)} \right]} \notag \\
\ge &\sum\limits_{k = 1}^K {{w_k}\frac{{\left( {1 - \eta } \right)}}{{\ln 2}}\left[ {\ln {{\left( {{\chi _k}} \right)}^{\left[ {{\rho ^{\left( i \right)}_{k}} - {\alpha _k}{{\hat \rho }^{\left( i \right)}_{k}}} \right]}} + {\delta ^{\left( i \right)}_{k}} - {\alpha _k}{{\hat \delta }^{\left( i \right)}_{k}}} \right]} \label{equal_OF},
\end{align}
where the equality holds only when $\chi_k = \chi_k^{\left(i \right)}$.

Next, we focus on the term consisting of variable $\chi_k$ in (\ref{equal_OF}), and solve the following subproblem in the $i$th iteration
\begin{subequations}
\setlength\abovedisplayskip{5pt}
\setlength\belowdisplayskip{5pt}
\label{equal_formualtion}
\begin{align}
\mathop {\max }\limits_{\left\{ {p_{m,k}^d} \right\},\left\{ {{\chi _k}} \right\}} & \prod\limits_{k = 1}^K {{\chi _k}^{{{\hat w}_k^{\left( i \right)}}}}  \label{equal_formualtion_a}\\
{\text{s.t.}}\;\;\;\;\;\;\;\; & {\rm{(\ref{MRC_optimization_sim_trans_b})}}, {\rm{ \left(\ref{MRC_optimization_sim_trans_c}\right)}}, {\rm{ \left(\ref{MRC_optimization_sim_c}\right)}} \label{equal_formualtion_b},
\end{align}
\end{subequations}
where ${{\hat w}^{\left( i \right)}_k}$ is equal to ${{\hat w}^{\left( i \right)}_k} = {w_k}\frac{{\left( {1 - \eta } \right)}}{{\ln 2}}\left( {{\rho ^{\left( i \right)}} - {\alpha _k}{{\hat \rho }^{\left( i \right)}}} \right)$.

Obviously, the problem in (\ref{equal_formualtion}) is not a GP problem as constraint (\ref{MRC_optimization_sim_trans_b}) is not a monomial function \cite{ref22}. To tackle this issue, considering the different expressions for various precoding schemes, we denote the numerator and the denominator of SINR $\hat \gamma_k$ as $\left(\theta_k\right)^2$ and $\varpi_k$, respectively, and then introduce a general theorem to approximate $\theta_k$ based on abovementioned three precoding schemes as a monomial function.

\begin{theorem}
\label{theta_T}
For any given $\hat p_{m,k}^d > 0$, $\theta_k$ is lower bounded by
\begin{equation}
	\setlength\abovedisplayskip{5pt}
	\setlength\belowdisplayskip{5pt}
	\begin{split}
		{\theta _k} &= {{\sum\limits_{m \in {{\cal M}_k}} {\sqrt {\left( {N - t_m} \right)p_{m,k}^d{\hat \lambda _{m,k}}} } } }  \\
		&\ge c_k \prod\limits_{m \in {\cal M}_k} {{{\left[ {\left( {N - {t_m}} \right)p_{m,k}^d{\hat \lambda _{m,k}}} \right]}^{{a_{m,k}}}}} \label{MRC_theta_LB} ,
	\end{split}
\end{equation}
where $\hat \lambda _{m,k}$ is equal to $\hat \lambda _{m,k} = \frac{KP_k^{{\rm max},p} \left(\beta_{m,k} \right)^2}{KP_k^{{\rm max},p}\beta_{m,k} + 1}$, $t_m$ is a constant that depends on the different precoding schemes, $a_{m,k}$ and $c_k$ are the coefficients. Specifically, $t_m$ is give by 
\begin{equation}
	\setlength\abovedisplayskip{5pt}
	\setlength\belowdisplayskip{5pt}
	\label{t_m}
	{t_m} = \left\{ {\begin{array}{*{20}{c}}
			{\begin{array}{*{20}{c}}
					{0,}\\
					{K,}\\
					{{\tau _m},}
			\end{array}}&{\begin{array}{*{20}{c}}
					{{\rm{MRT}}}\\
					{{\rm{FZF}}}\\
					{{\rm{LZF}}}
			\end{array}}
	\end{array}} \right..
\end{equation}
The coefficients $a_{m,k}$ and $c_k$ are given by
\begin{equation}
\setlength\abovedisplayskip{5pt}
\setlength\belowdisplayskip{5pt}
\label{a_k}
{a_{m,k}} = \frac{{\sqrt {\left( {N - {t_m}} \right)\hat p_{m,k}^d{{\hat \lambda }_{m,k}}} }}{{{2{\hat \theta }_k}}},
\end{equation}
and
\begin{equation}
\setlength\abovedisplayskip{5pt}
\setlength\belowdisplayskip{5pt}
\label{c_k}
{c_k} = \frac{{{{\hat \theta} _k}}}{{\prod\limits_{m \in {\cal M}_k} {{{\left[ {\left( {N - {t_m}} \right){\hat p}_{m,k}^d{\hat \lambda _{m,k}}} \right]}^{{a_{m,k}}}}} }},
\end{equation}
where $\hat \theta _k$ is obtained by using $p^d_{m,k} = \hat p^d_{m,k}$. Besides, it is obvious that the inequality in (\ref{MRC_theta_LB}) holds with equality when $p^d_{m,k} = \hat p^d_{m,k}$.

{\emph{Proof}}: Please refer to Appendix \ref{proof_theta_T}. $\hfill\blacksquare$

\end{theorem}

By using Theorem \ref{theta_T}, similar to the objection function in (\ref{MRC_optimization_sim_trans_a}), we can approximate the numerator $(\theta_k)^2$ in an iterative manner. Specifically, $c_k^{\left(i\right)}$ and $a_{m,k}^{\left(i\right)}$ are obtained based on (\ref{a_k}) and (\ref{c_k}) by using ${\hat p}_{m,k}^d = p_{m,k}^{d,(i)}$, and $\theta_k$ can be lower bounded by
\begin{equation}
\setlength\abovedisplayskip{5pt}
\setlength\belowdisplayskip{5pt}
{\theta _k} \ge c_k^{\left(i\right)} \prod\limits_{m \in {\cal M}_k} {{{\left[ {\left( {N - {t_m}} \right)p_{m,k}^d{\hat \lambda _{m,k}}} \right]}^{{a_{m,k}^{\left(i\right)}}}}} \label{MRC_theta_LB_i}.
\end{equation}

Based on the abovementioned simplifications and approximations, the problem is transformed into the following GP problem
\begin{subequations}
	\setlength\abovedisplayskip{5pt}
	\setlength\belowdisplayskip{5pt}
	\label{MRC_optimization_final}
	\begin{align}
		\mathop {\max }\limits_{\left\{ {p_{m,k}^d} \right\},\left\{ {{\chi _k}} \right\}} & \prod\limits_{k = 1}^K {{\chi _k}^{{{\hat w}_k^{\left( i \right)}}}} \label{MRC_optimization_final_a} \\
		\text{s.t.} \;\;\;\;\; & \left(c_k^{\left(i\right)}\right)^2 \prod\limits_{m \in {\cal M}_k} {{{\left[ {\left( {N - {t_m}} \right)p_{m,k}^d{\hat \lambda _{m,k}}} \right]}^{{2 a_{m,k}^{\left(i\right)}}}}} \notag \\
		& \ge {\chi _k} \varpi_k ,\forall k,  \label{MRC_optimization_final_b}\\
		& {\rm{ \left(\ref{MRC_optimization_sim_trans_c}\right)}}, {\rm{ \left(\ref{MRC_optimization_sim_c}\right)}}. \label{MRC_optimization_final_d}
	\end{align}
\end{subequations}

To run the iterative algorithm, it is necessary to find a feasible initial solution. To deal with this issue, we construct an alternative optimization problem by introducing an auxiliary variable $\varphi$, which is given by
\begin{subequations}
	\setlength\abovedisplayskip{5pt}
	\setlength\belowdisplayskip{5pt}
	\label{feasible_region}
	\begin{align}
		\mathop {\max } \limits_{\varphi ,\left\{ {p_{m,k}^d} \right\}}  & \varphi \label{feasible_region_a} \\
		{\text{s.t.}}\;\;\;\;\;  & \left(c_k^{\left(i\right)}\right)^2 \prod\limits_{m \in {\cal M}_k} {{{\left[ {\left( {N - {t_m}} \right)p_{m,k}^d{\hat \lambda _{m,k}}} \right]}^{{2 a_{m,k}^{\left(i\right)}}}}} \notag \\
		&\ge  \frac{\varphi }{{f_k^{ - 1}\left( {\frac{{R_k^{{\rm{req}}}\ln 2}}{{1 - \eta }}} \right)}} \varpi_k, \label{feasible_region_b} \\
		& ({\rm{\ref{MRC_optimization_sim_c}}}) \label{feasible_region_c}.
	\end{align}
\end{subequations}

Obviously, Problem (\ref{feasible_region}) is also a GP problem, and is always feasible. Besides, Problem (\ref{MRC_optimization_final}) is feasible only when $\varphi$ is no smaller than 1. \textcolor{black}{Furthermore, we define an error tolerance $\xi$ to guarantee that the transmission power converges to the optimal solutions}. Based on the abovementioned discussions, Algorithm 1 is provided to maximize the weighted sum rate.
\begin{algorithm}[t]
\caption{Iterative Algorithm for Solving Maximum Weighted Sum Rate}
\begin{algorithmic}[1]
\label{MRC_algorithm}
\STATE Initialize iteration number $i = 1$, and error tolerance $\zeta = 0.01$;
\STATE Initialize the pilot power $\left\{p_k^p = P^{{\rm max},p}_k,\forall k\right\}$, calculate transmission power $\left\{ p^{d,\left( 1\right)}_{m,k},\forall m, k \right\}$ by solving Problem (\ref{feasible_region}), obtain SINR $\left \{\chi _k^{\left( 1 \right)},\forall k\right\}$ and the weighted sum rate in (\ref{MRC_optimization_a}) denoted as ${\rm{Obj}}^{\left(1\right)}$. Set ${\rm{Obj}}^{\left(0\right)} = {\rm{Obj}}^{\left(1\right)} \zeta$;
\WHILE {${{\left( {{\rm{Ob}}{{\rm{j}}^{\left( i \right)}} - {\rm{Ob}}{{\rm{j}}^{\left( {i - 1} \right)}}} \right)} \mathord{\left/
 {\vphantom {{\left( {{\rm{Ob}}{{\rm{j}}^{\left( i \right)}} - {\rm{Ob}}{{\rm{j}}^{\left( {i - 1} \right)}}} \right)} {{\rm{Ob}}{{\rm{j}}^{\left( {i - 1} \right)}}}}} \right.
 \kern-\nulldelimiterspace} {{\rm{Ob}}{{\rm{j}}^{\left( {i - 1} \right)}}}} \ge \zeta$}
\STATE Update $\left \{ {{\hat w}^{\left( i \right)}_k},c^{\left(i\right)}_k, a^{\left(i\right)}_{m,k},\forall m,k \right\}$;
\STATE Update $i = i+1$, solve Problem (\ref{MRC_optimization_final}) by using the CVX package to obtain $\left \{p_{m,k}^{d,\left( i \right)} ,\forall m,k\right\}$, calculate SINR $\left \{\chi _k^{\left( i \right)},\forall k\right\}$ and then obtain the weighted sum rate, denoted as ${\rm{Obj}}^{\left(i\right)}$;
\ENDWHILE
\end{algorithmic}
\end{algorithm}

\subsection{Algorithm Analysis}
1) \emph{Feasibility Analysis}:
For Algorithm \ref{MRC_algorithm}, we need to check whether constraint (\ref{MRC_optimization_final_b}) in the $i$th iteration holds or not in the $(i+1)$th iteration as only constraint (\ref{MRC_optimization_final_b}) is approximated in an iterative manner. Constraint (\ref{MRC_optimization_final_b}) in the $i$th iteration is given by
\begin{equation}
\setlength\abovedisplayskip{5pt}
\setlength\belowdisplayskip{5pt}
\label{MRC_sinr_appro_n}
\left(c_k^{\left(i - 1\right)}\right)^2 \prod\limits_{m \in {\cal M}_k} {{{\left[ {\left( {N - {t_m}} \right)p_{m,k}^{d,{\left(i\right)}}{\hat \lambda _{m,k}}} \right]}^{{2a_{m,k}^{\left(i - 1\right)}}}}} \ge {\chi _k^{\left(i\right)}} \varpi_k^{\left(i\right)},
\end{equation}
where $\left\{ \chi _k^{\left( i \right)}, p_{m,k}^{d,\left( i \right)}, \forall m,k\right\}$ is the optimal solution in the $i$th iteration, and $\varpi_k^{\left(i\right)}$ is obtained by using $p_{m,k}^d = p_{m,k}^{d,\left( i \right)}$.

Using Theorem \ref{theta_T} and ({\ref{MRC_theta_LB_i}}), we have
\begin{equation}
\setlength\abovedisplayskip{5pt}
\setlength\belowdisplayskip{5pt}
\label{MRC_sinr_appro_n1}
\begin{aligned}
 & c_k^{\left(i\right)} \prod\limits_{m \in {\cal M}_k} {{{\left[ {\left( {N - {t_m}} \right)p_{m,k}^{d,{\left(i\right)}}{\hat \lambda _{m,k}}} \right]}^{{a_{m,k}^{\left(i\right)}}}}} = \theta_k^{\left(i \right)}\\
& \ge c_k^{\left(i - 1\right)} \prod\limits_{m \in {\cal M}_k} {{{\left[ {\left( {N - {t_m}} \right)p_{m,k}^{d,{\left(i\right)}}{\hat \lambda _{m,k}}} \right]}^{{a_{m,k}^{\left(i - 1\right)}}}}}.
\end{aligned}
\end{equation}
Then, by combining (\ref{MRC_sinr_appro_n}) with (\ref{MRC_sinr_appro_n1}), we have
\begin{equation}
\setlength\abovedisplayskip{5pt}
\setlength\belowdisplayskip{5pt}
\label{MRC_sinr_appro_n2}
\left(c_k^{\left(i\right)}\right)^2 \prod\limits_{m \in {\cal M}_k} {{{\left[ {\left( {N - {t_m}} \right)p_{m,k}^{d,{\left(i\right)}}{\hat \lambda _{m,k}}} \right]}^{{2a_{m,k}^{\left(i\right)}}}}} \ge {\chi _k^{\left(i\right)}} \varpi_k^{\left(i\right)}.
\end{equation}
Obviously, the solution is also feasible in the $(i+1)$th iteration.

2) \emph{Convergence Analysis}:
We prove that our algorithm can converge to a locally optimal solution. Denote ${\rm{Obj}}^{\left(i \right)}$ as the weighted sum rate in the $i$th iteration. Since the solution in the $i$th iteration is also feasible in the $(i+1)$th iteration, we have
\begin{equation}
\setlength\abovedisplayskip{5pt}
\setlength\belowdisplayskip{5pt}
 \label{MRC_Cov1}
\begin{split}
 &\sum\limits_{k = 1}^K {{w_k}\frac{{\left( {1 - \eta } \right)}}{{\ln 2}} \left[ {\ln {{\left( {{\chi ^{\left( i + 1 \right)}_k}} \right)}^{\left[ {{\rho ^{\left( i \right)}_{k}} - {\alpha _k}{{\hat \rho }^{\left( i \right)}_{k}}} \right]}} + {\delta ^{\left( i \right)}_{k}} - {\alpha _k}{{\hat \delta }^{\left( i \right)}_{k}}} \right]} \\
\ge & \sum\limits_{k = 1}^K {{w_k}\frac{{\left( {1 - \eta } \right)}}{{\ln 2}}  \left[ {\ln {{\left( {{\chi ^{\left( i \right)}_k}} \right)}^{\left[ {{\rho ^{\left( i \right)}_{k}} - {\alpha _k}{{\hat \rho }^{\left( i \right)}_{k}}} \right]}} + {\delta ^{\left( i \right)}_{k}} - {\alpha _k}{{\hat \delta }^{\left( i \right)}_{k}}} \right]} \\
 =  & {\rm{Obj}}^{\left(i\right)},
 \end{split}
\end{equation}
where $\left\{\chi_k^{\left(i+1\right)},\forall k\right\}$ is the optimal solution to Problem (\ref{MRC_optimization_final}) in the $\left(i+1\right)$th iteration.

Substituting ${\chi _k }= {\chi ^{\left( i + 1 \right)}_k} $ into the inequality in (\ref{equal_OF}), we have
\begin{equation}
	\setlength\abovedisplayskip{5pt}
	\setlength\belowdisplayskip{5pt}
	\label{MRC_inequality}
	\begin{split}
		& \sum\limits_{k = 1}^K {{w_k}\frac{{\left( {1 - \eta } \right)}}{{\ln 2}}\left[ {\ln \left( {1 + {\chi ^{\left( i + 1 \right)}_k} } \right) - {\alpha _k}G\left( {{\chi ^{\left( i + 1 \right)}_k} } \right)} \right]}  \\
		\ge \!  & \sum\limits_{k = 1}^K {{w_k}\frac{{\left( {1 - \eta } \right)}}{{\ln 2}}} \times \\
		& \left[ {\ln {{\left( {{\chi ^{\left( i + 1 \right)}_k}} \right)}^{\left[ {{\rho ^{\left( i + 1 \right)}_{k}}  \! -  \! {\alpha _k}{{\hat \rho }^{\left( i + 1\right)}_{k}}} \right]}}  \!+ \! {\delta ^{\left( i + 1 \right)}_{k}}  \!- \! {\alpha _k}{{\hat \delta }^{\left( i + 1\right)}_{k}}} \right]  \\
		\ge &\sum\limits_{k = 1}^K {{w_k}\frac{{\left( {1 - \eta } \right)}}{{\ln 2}}   \left[ {\ln {{\left( {{\chi ^{\left( i + 1 \right)}_k}} \right)}^{\left[ {{\rho ^{\left( i \right)}_{k}} - {\alpha _k}{{\hat \rho }^{\left( i \right)}_{k}}} \right]}} + {\delta ^{\left( i \right)}_{k}} - {\alpha _k}{{\hat \delta }^{\left( i \right)}_{k}}} \right]} .
	\end{split}
\end{equation}

Then, by combining (\ref{MRC_Cov1}) with (\ref{MRC_inequality}), we have
\begin{equation}
	\setlength\abovedisplayskip{5pt}
	\setlength\belowdisplayskip{5pt}
	\label{MRC_Cov2}
	\begin{split}
		& {\rm{Obj}}^{\left(i+1\right)} \\
		& =  \sum\limits_{k = 1}^K {{w_k}\frac{{\left( {1 - \eta } \right)}}{{\ln 2}} \left[ {\ln \left( {1 + {\chi ^{\left( i + 1 \right)}_k}} \right) - {\alpha _k}G\left( {{\chi ^{\left( i + 1 \right)}_k}} \right)} \right]} \\
		& \ge \sum\limits_{k = 1}^K {{w_k}\frac{{\left( {1 - \eta } \right)}}{{\ln 2}}  \left[ {\ln {{\left( {{\chi ^{\left( i + 1 \right)}_k}} \right)}^{\left[ {{\rho ^{\left( i \right)}_{k}} - {\alpha _k}{{\hat \rho }^{\left( i \right)}_{k}}} \right]}} + {\delta ^{\left( i \right)}_{k}} - {\alpha _k}{{\hat \delta }^{\left( i \right)}_{k}}} \right]} \\
		& \ge   {\rm{Obj}}^{\left(i\right)}  .
	\end{split}
\end{equation}
Therefore, the convergence of Algorithm \ref{MRC_algorithm} is verified. Besides, we can prove that Algorithm \ref{MRC_algorithm} can converge to the Karush-Kuhn-Tucker (KKT) point of Problem (\ref{MRC_optimization}) for the abovementioned precoding schemes by using the similar proof as in Appendix B in \cite{ref21}. 

3) \emph{Complexity Analysis}:
\textcolor{black}{The complexity of Algorithm 1 depends on the number of iterations and complexity of each iteration. Specifically, the main complexity of each iteration in Algorithm 1 lies in solving Problem (\ref{MRC_optimization_final}) which includes $(M + 1)K$ variables and $(2K + M)$ constraints. Based on \cite{ref22}, the computational complexity of this algorithm is on the order of ${\mathcal{O}}(N_{iter} \times \max\{[(M + 1)K]^{3},(2K + M)[(M + 1)K]^{2}, N_{cost}\})$, where $N_{iter}$ is the number of iterations and $N_{cost}$ is the computational complexity of calculating the first-order and second-order derivatives of the objective function and constraint functions of Problem (\ref{MRC_optimization_final}) \cite{van2018joint}. Furthermore, our simulation results demonstrate that Algorithm 1 can converge to the locally optimal solution with fewer iterations.}

\section{Simulation Results}
The performance of the proposed algorithms are numerically evaluated and discussed in this section. We first introduce the simulation setup and the related simulation parameters.
\subsection{Simulation Scenario}
The smart factory is assumed to be located in a $\rm D$ $\times$ $\rm D$ square. \textcolor{black}{In contrast to the wraparound deployment in \cite{zhang2021improving,wang2022uplink}, we uniformly deploy $M$ APs at constellation points to provide uniform service for the devices.} The large-scale fading coefficient model is adopted \cite{2017small}, which is given by
\begin{equation}
	\setlength\abovedisplayskip{5pt}
	\setlength\belowdisplayskip{5pt}
	\label{channel_model}
	{\rm{P}}{{\rm{L}}_{m,k}} = \left\{ {\begin{array}{*{20}{l}}
			{\begin{array}{*{20}{l}}
					{L_{\rm{loss}} \!+\! 35{{\log }_{10}}\left( {{d_{m,k}}} \right),{{d_{m,k}} \!>\! {d_1}}, }\\
					{L_{\rm{loss}} \!+\! 15{{\log }_{10}}\left( {{d_1}} \right) \!+\! 20{{\log }_{10}}\left( {{d_0}} \right),{{d_{m,k}} \!\le\! {d_0}},} \\
					{L_{\rm{loss}} \!+\! 15{{\log }_{10}}\left( {{d_1}} \right) \!+\! 20{{\log }_{10}}\left( {{d_{m,k}}} \right), \rm{other},}\\
			\end{array}}
	\end{array}} \right.
\end{equation}
where \textcolor{black}{$d_{m,k} \left(\rm{km}\right)$} is the distance between the $m$th AP and the $k$th device, and $\textcolor{black}{L_{\rm{loss}} \left(\rm{dB}\right)}$ is a constant factor that depends on the carrier frequency $f\left(\rm{MHz}\right)$, the heights of the APs $h_{{\rm{AP}}} \left(\rm{m}\right)$ and devices $h_u \left(\rm{m}\right)$. Specifically, $L_{\rm {loss}}$ is given by
\begin{equation}
\setlength\abovedisplayskip{5pt}
\setlength\belowdisplayskip{5pt}
\label{path_loss_L}
\begin{split}
L_{\rm{loss}} &= 46.3 + 33.9{\log _{10}}\left( f \right) - 13.82{\log _{10}}\left( {{h_{{\rm{AP}}}}} \right) \\
 & - \left( {1.1{{\log }_{10}}\left( f \right) - 0.7} \right){h_u} + \left( {1.56{{\log }_{10}}\left( f \right) - 0.8} \right).
\end{split}
\end{equation}
\textcolor{black}{Besides, for the small-scale fading, it is generally modeled as Rayleigh fading with zero mean and unit variance.}
\textcolor{black}{The corresponding normalized pilot power $p_k^p$ and transmission power $p^d_{m,k}$ can be computed through dividing these powers by the noise power, which is given by}
\begin{equation}
\setlength\abovedisplayskip{5pt}
\setlength\belowdisplayskip{5pt}
\label{noise_power}
P_n = B \times {k_B} \times {T_0} \times  10^{\frac{N_{\rm{dB}}}{10}} \left( {\rm{W}} \right),
\end{equation}
where $k_B = 1.381 \times 10^{-23}$ (Joule per Kelvin) is the Boltzmann constant, and ${T_0} = 290$ (Kelvin) is the noise temperature. The weights for all the devices are randomly generated within [0,1]. \textcolor{black}{Unless otherwise specified, the simulation parameters are similar to those in \cite{ref16,ref18} and summarized in Table I}. \textcolor{black}{More importantly, we fix the total number of antennas in this smart factory to investigate the deployment of APs.} In other words, if each AP is equipped with more antennas, this area will deploy less APs.

As mentioned before, the $k$th device is served by the set of APs ${\cal M}_k$. Specifically, it is assumed that the large-scale fading factors are known at the $m$th AP, and then the large-scale fading factors $\left\lbrace {\beta_{1,k},\beta_{2,k},\cdot \cdot \cdot, \beta_{M ,k}}\right\rbrace $ are sorted in a descending order. Finally, the large-scale fading factors are selected in turn until satisfying the following condition 
\begin{equation}
\setlength\abovedisplayskip{5pt}
\setlength\belowdisplayskip{5pt}
\label{AP_selection}
\frac{\sum \nolimits_{m \in {\mathcal{M}}_k} {\beta_{m,k}}}{\sum\nolimits_{m = 1}^M {{\beta _{m,k}}}} \ge  T_h,
\end{equation}
where $T_h$ is the threshold. For the set of devices served by the $m$th AP, by checking whether the $m$th AP belongs to the set of $\mathcal{M}_k$, $k = 1,2,\cdot \cdot \cdot, K$, we can obtain $\mathcal{U}_m$.

\begin{table}[t]
	\small
	\caption{Simulation Parameters}
	\centering
	\begin{tabular}{|c|c|}\hline
		Parameters Setting & Value \\ \hline
		Carrier frequency ($f$) & 2.1 GHz \\ Bandwidth ($B$) & 10 MHz \\
		Transmission duration ($T_B$) & 0.05 ms \\ Blocklength ($L = B T_B$) & 500 \\
		Height of APs (${h_{{\rm{AP}}}}$)& 15 m \\ Height of devices ($h_u$) & 1.6 m \\
		Noise figure ($N_{\rm{dB}}$) & 9 dB \\  Number of devices ($K$) & 10 \\
		Required data rate ($R_{\rm{req}}$) & 0.5 bit/s/Hz \\ Decoding error probability ${{\varepsilon _k}}$ & $10^{-7}$ \\
		Size of square ($\rm D$) & 1000 m \\ Pilot power $P_k^{\rm{max}}, \forall k$ & $100$ mW \\
		$d_{0}$ & 10 m \\ $d_{1}$ & 50 m \\ \hline
	\end{tabular}
	\label{tab: Margin_settings}
\end{table}

\subsection{Properties of the Proposed Algorithm}
In this subsection, we first check the gap between the LB data rate and the ergodic data rate, illustrate the convergence behavior of the proposed algorithm, and then investigate the impact of threshold on the system performance.
\begin{figure*}[t]
	\begin{minipage}[t]{0.32\linewidth}
		{\includegraphics[width=2.25in]{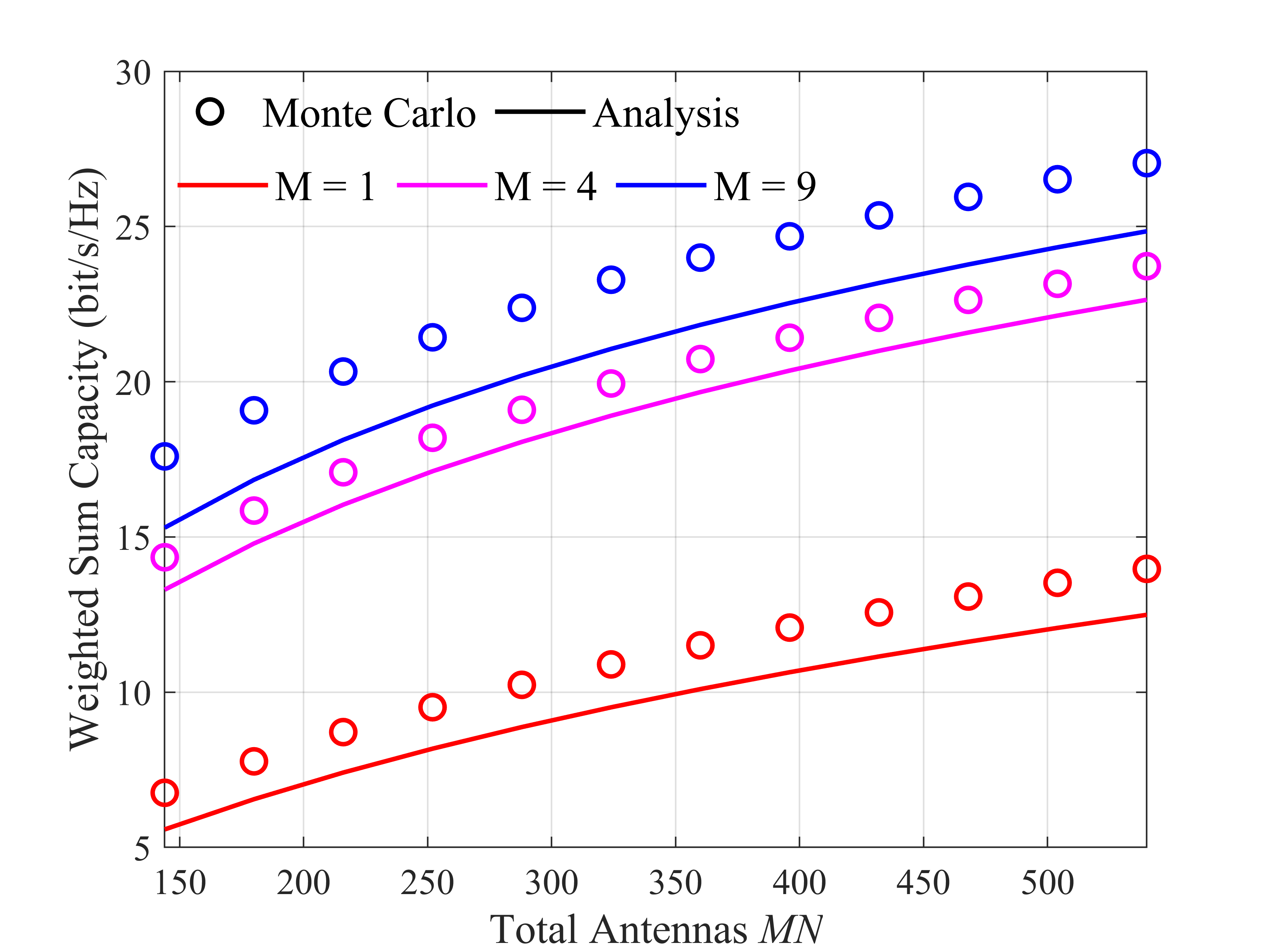}%
			\caption*{(a) MRT}
			\label{fig:MRC_Tightness}}
	\end{minipage}
	\hfill
	\begin{minipage}[t]{0.32\linewidth}
		{\includegraphics[width=2.25in]{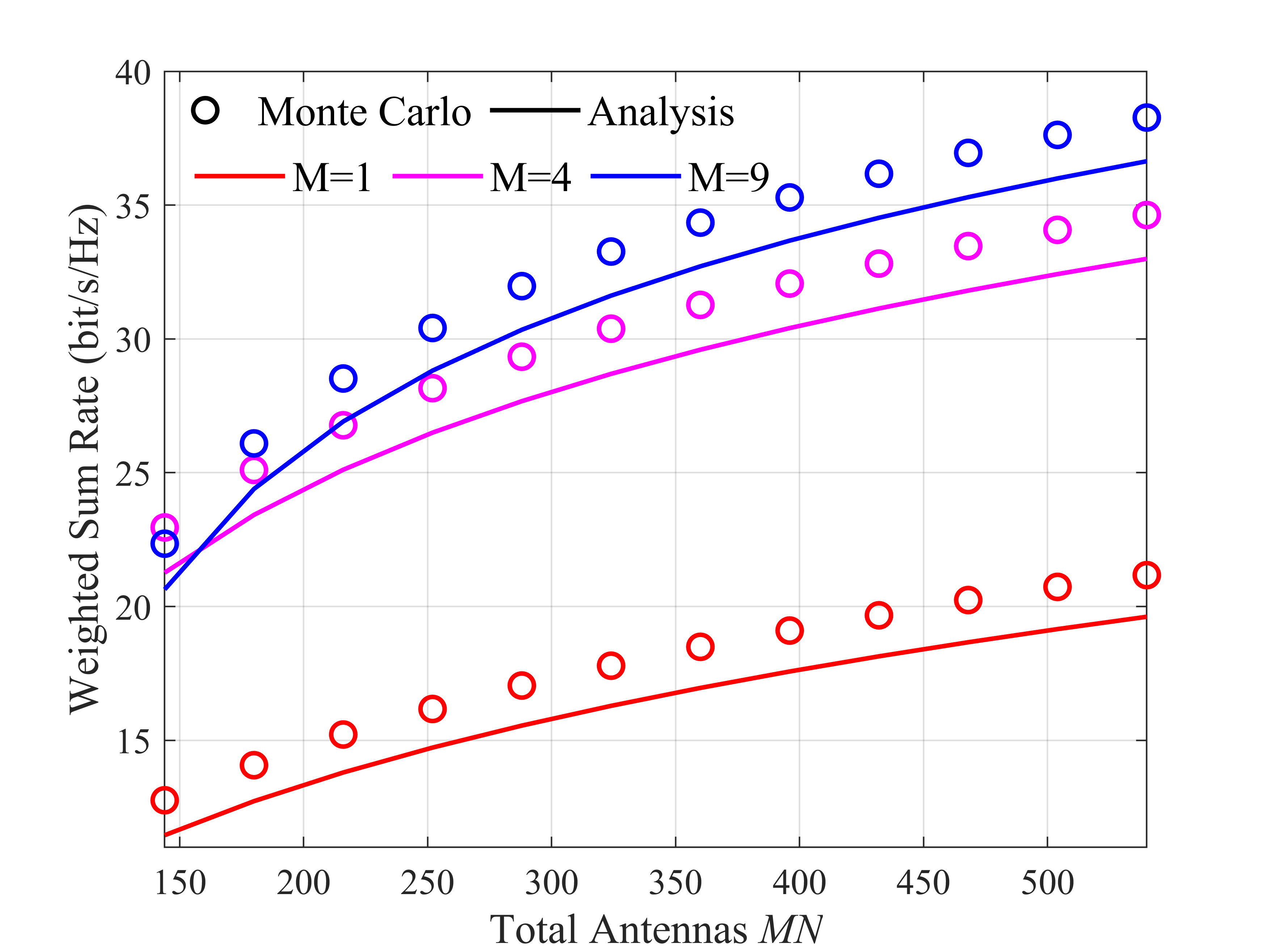}%
			\label{fig:FZF_Tightness}}
		\caption*{(b) FZF}
	\end{minipage}
	\hfill
	\begin{minipage}[t]{0.32\linewidth}
		{\includegraphics[width=2.25in]{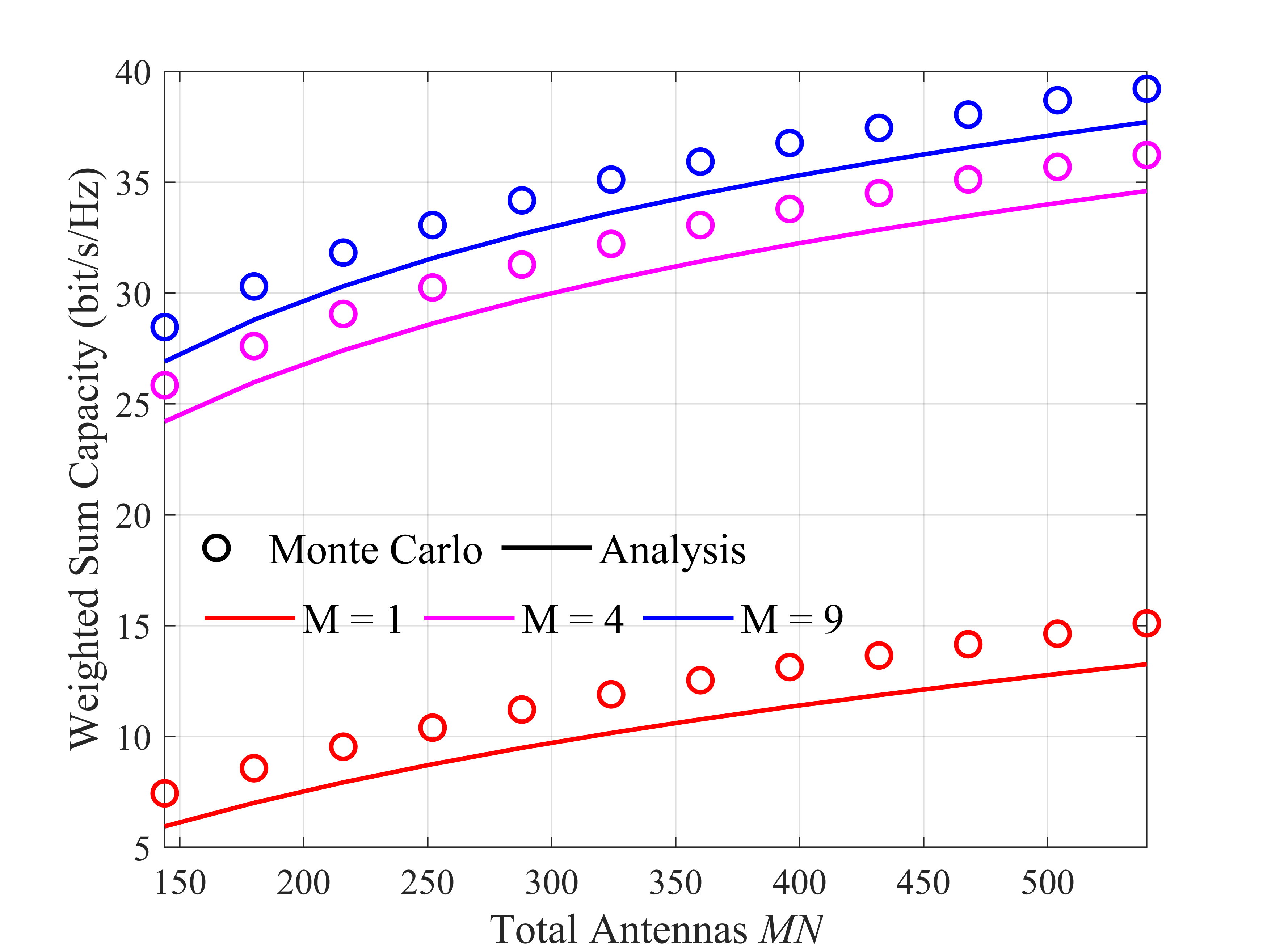}%
			\label{fig:LZF_Tightness}}
		\caption*{(c) LZF}
	\end{minipage}
	\caption{Weighted Sum Rate V.S. The Number of Total Antennas under various numbers of APs.}
	\label{fig_Tightness}
\end{figure*}

1) \emph{Tightness}: The simulation results are obtained through the Monte-Carlo simulation by averaging over $10^4$ random channel generations with $T_h = 0.9$ and $p_{m,k}^d = 0.1$ $W$, $\forall m,k$. As can be seen from Fig. \ref{fig_Tightness}, the derived LB data rate is close to the ergodic rate for any system parameters, which confirms that the LB data rate is suitable and reasonable for power allocation.

2) \emph{Convergence}:
We investigate the convergence behavior of the proposed algorithm  with $MN = 144$ in Fig. \ref{fig_Convergence}. For given any transmission power $P_m$ and threshold $T_h$, the system performance for three precoding schemes can converge to the locally optimal solution within only 2 or 3 iterations, which demonstrates the rapid convergence of the proposed algorithm.
\begin{figure*}[t]
\begin{minipage}[t]{0.32\linewidth}
{\includegraphics[width=2.25in]{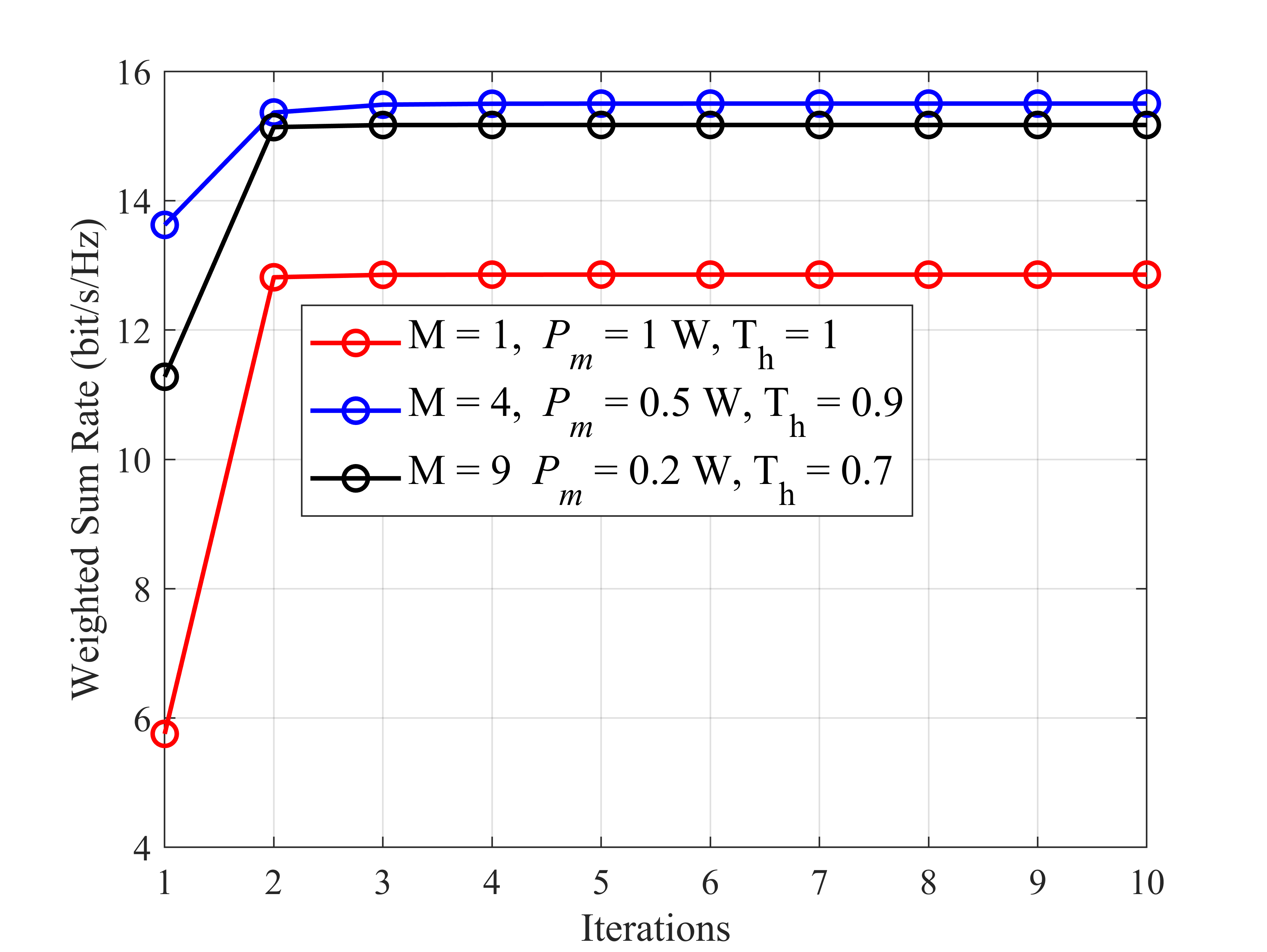}%
\caption*{(a) MRT}
\label{fig:MRC_Convergence}}
\end{minipage}
\hfill
\begin{minipage}[t]{0.32\linewidth}
{\includegraphics[width=2.25in]{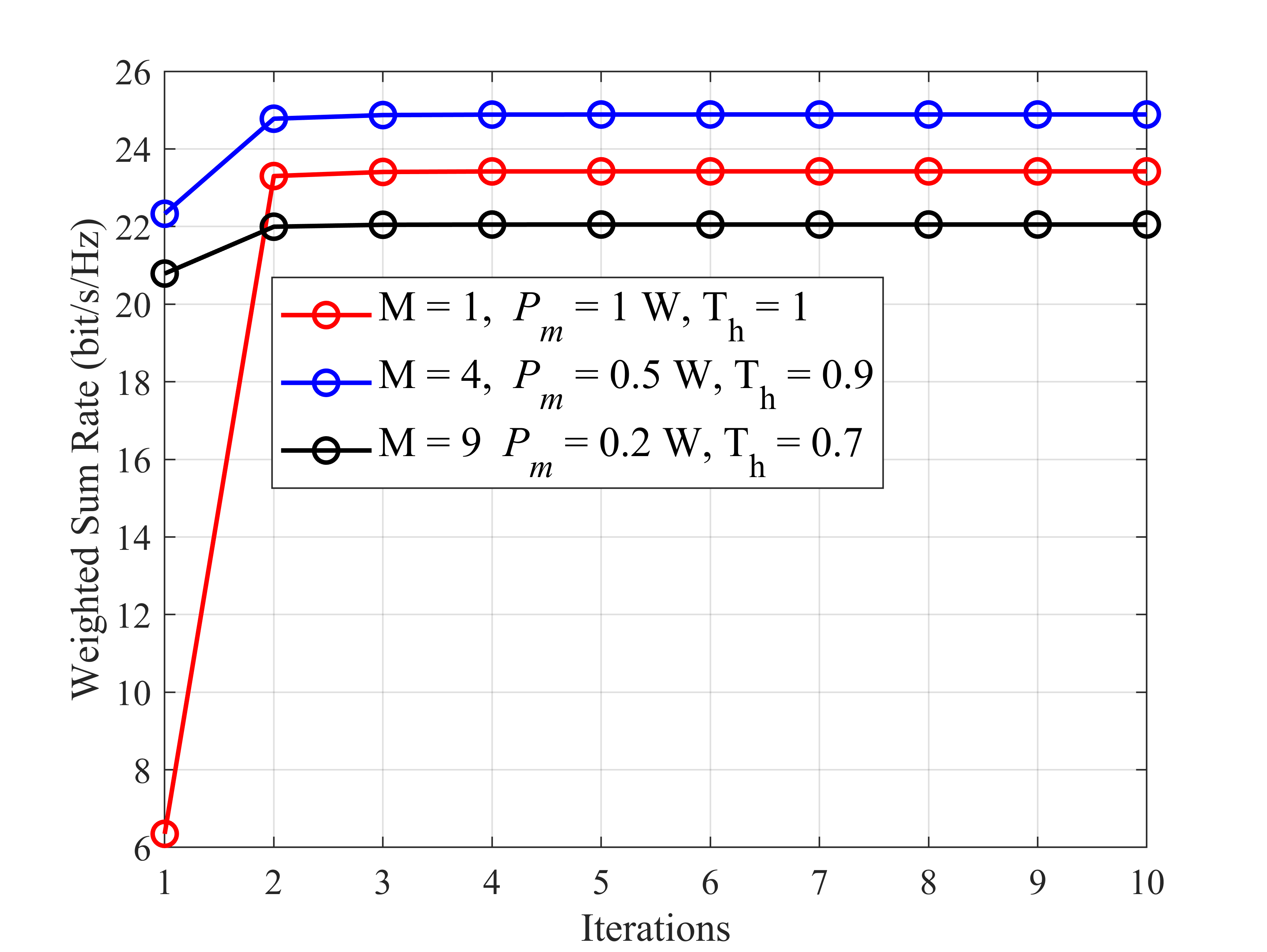}%
\label{fig:FZF_Convergence}}
\caption*{(b) FZF}
\end{minipage}
\hfill
\begin{minipage}[t]{0.32\linewidth}
{\includegraphics[width=2.25in]{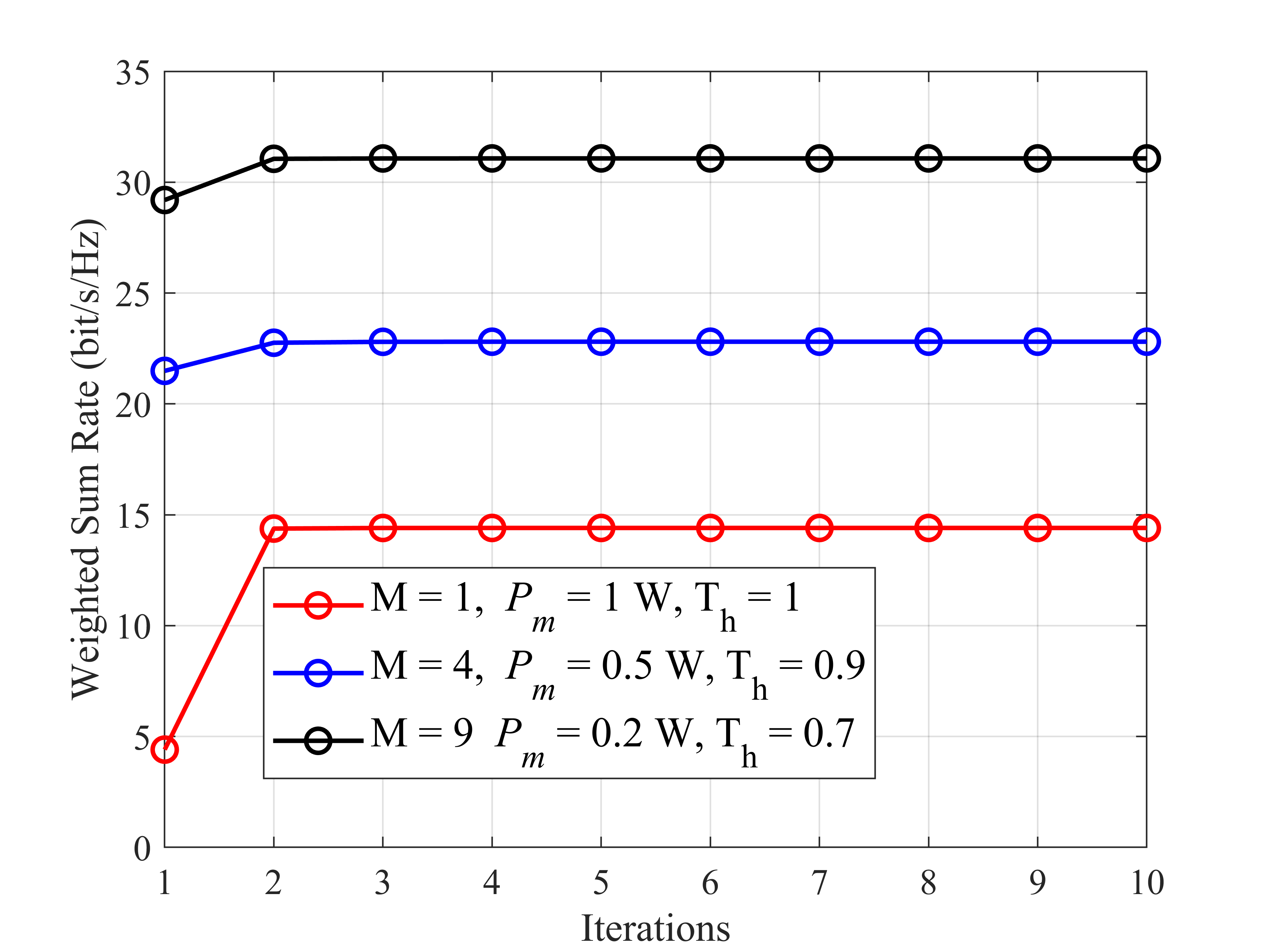}%
\label{fig:LZF_Convergence}}
\caption*{(c) LZF}
\end{minipage}
\caption{Convergence of proposed algorithm for different precoding schemes.}
\label{fig_Convergence}
\end{figure*}

3) \emph{Threshold}: The performance of the proposed algorithm is obtained by averaging $100$ random devices' locations and the system performance is set to zero if any devices cannot satisfy the data requirements. Fig. \ref{fig_Threshold} shows the system performance versus different thresholds with $P_m = 1$ $W$, $\forall m$. Obviously, it is observed that the optimal value of $T_h$ is $1$ for the MRT and FZF schemes and $0.95$ for the LZF precoder. This is due to the fact that selecting more APs to provide service for devices will consume the degrees of freedom for the LZF precoding scheme, leading to performance degradation. Here, we set $T_h = 0.95$ for all the following simulations, to achieve a good tradeoff between system performance and computational complexity.
\begin{figure*}[t]
\begin{minipage}[t]{0.32\linewidth}
{\includegraphics[width=2.25in]{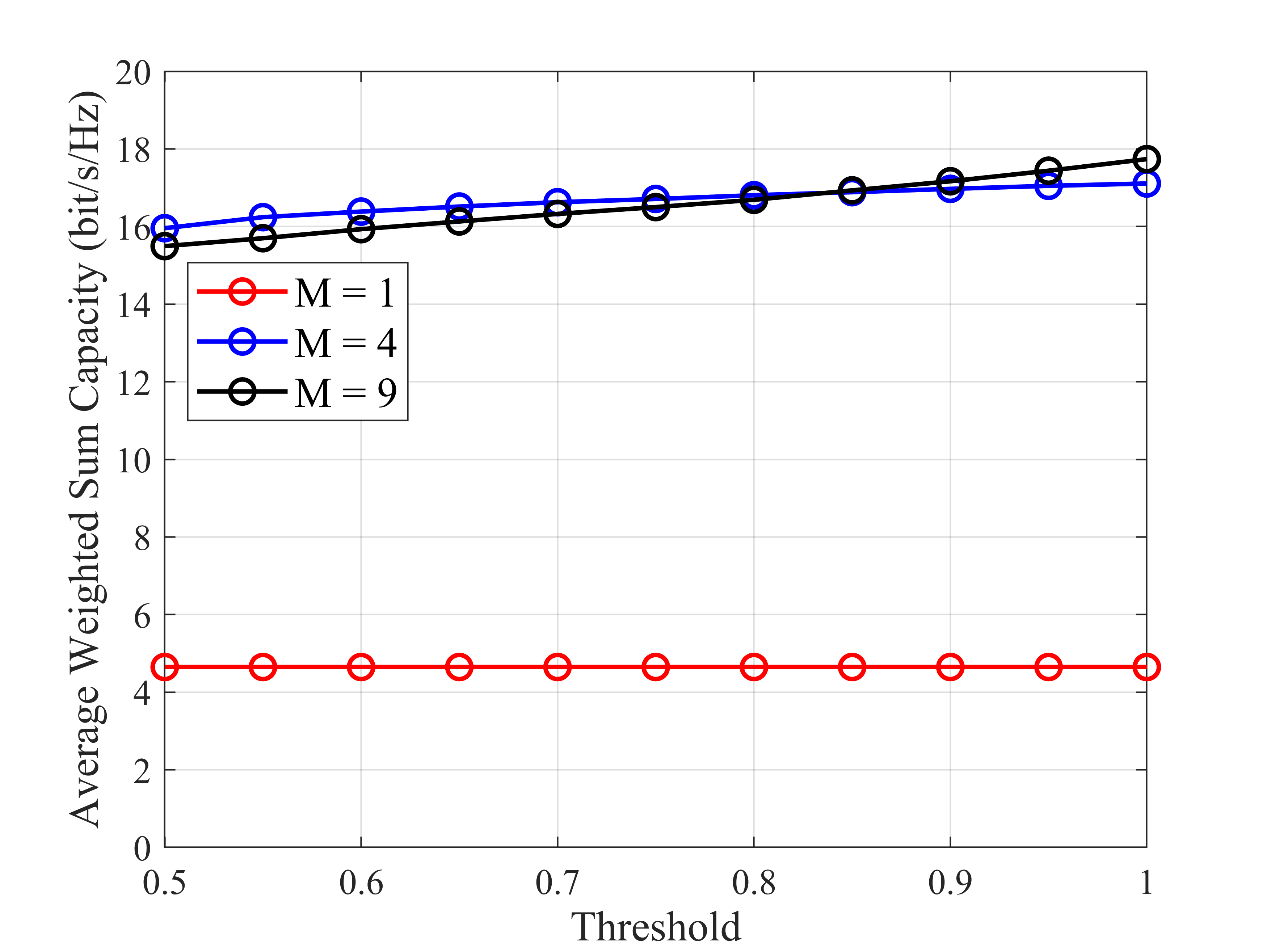}%
\caption*{(a) MRT}
\label{fig:MRC_Threshold}}
\end{minipage}
\hfill
\begin{minipage}[t]{0.32\linewidth}
{\includegraphics[width=2.25in]{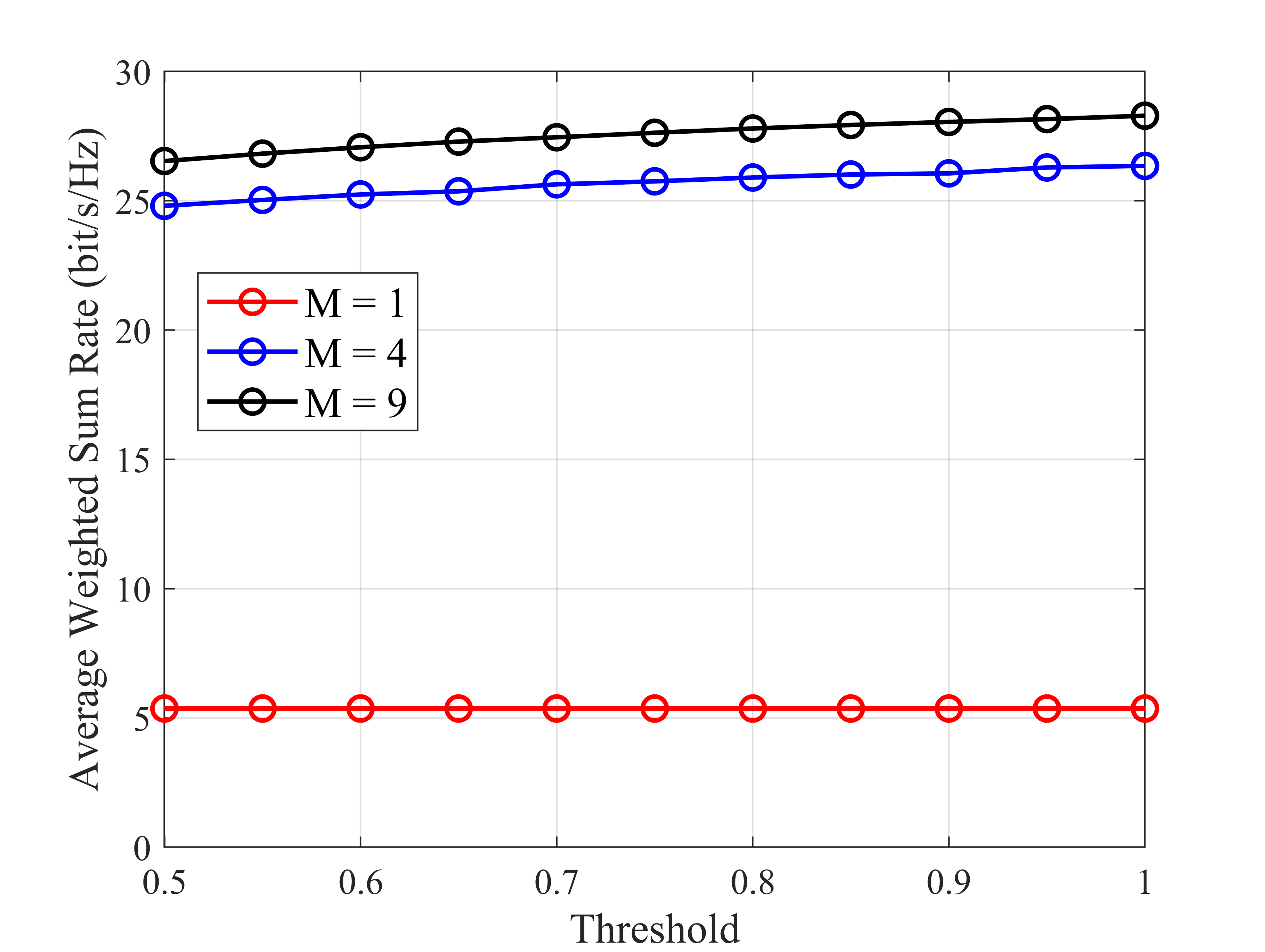}%
\label{fig:FZF_Threshold}}
\caption*{(b) FZF}
\end{minipage}
\hfill
\begin{minipage}[t]{0.32\linewidth}
{\includegraphics[width=2.25in]{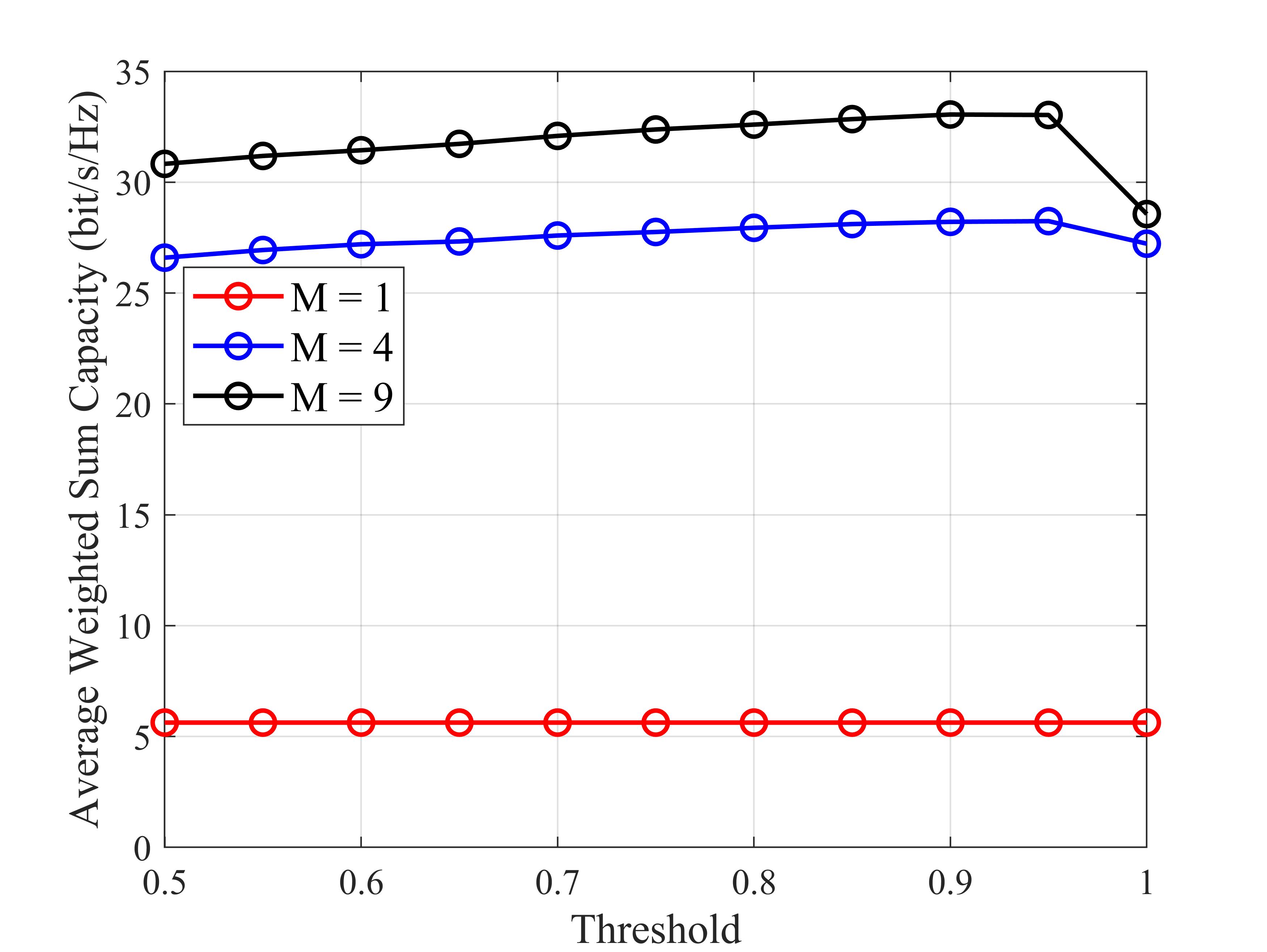}%
\label{fig:LZF_Threshold}}
\caption*{(c) LZF}
\end{minipage}
\caption{Performance of the proposed algorithm versus threshold for different precoding schemes.}
\label{fig_Threshold}
\end{figure*}

\subsection{Effect of pilot power}
In this subsection, we investigate how pilot power affects the system performance. Fig. \ref{fig_Pilot_Power} depicts the average weighted sum rate versus the pilot power with $MN = 144$ and $P_m = 1$ $W$, $\forall m$, by averaging $100$ random devices' locations. As can be seen from Fig. 5, the average weighted sum rate increases with the pilot power for any cases, which demonstrates that more accurate channel estimation is beneficial for enhancing the system performance. More importantly, we find an interesting phenomenon that the CF mMIMO (e.g., $M \ge 4$) significantly outperforms the centralized mMIMO system (e.g., $M = 1$) when the pilot power is low. This is attributed to the fact that the devices are closer to the APs in CF mMIMO systems than in centralized mMIMO systems, hence less pilot power is required to satisfy the requirements of DEP and data rate.
\begin{figure*}[t]
\begin{minipage}[t]{0.32\linewidth}
{\includegraphics[width=2.25in]{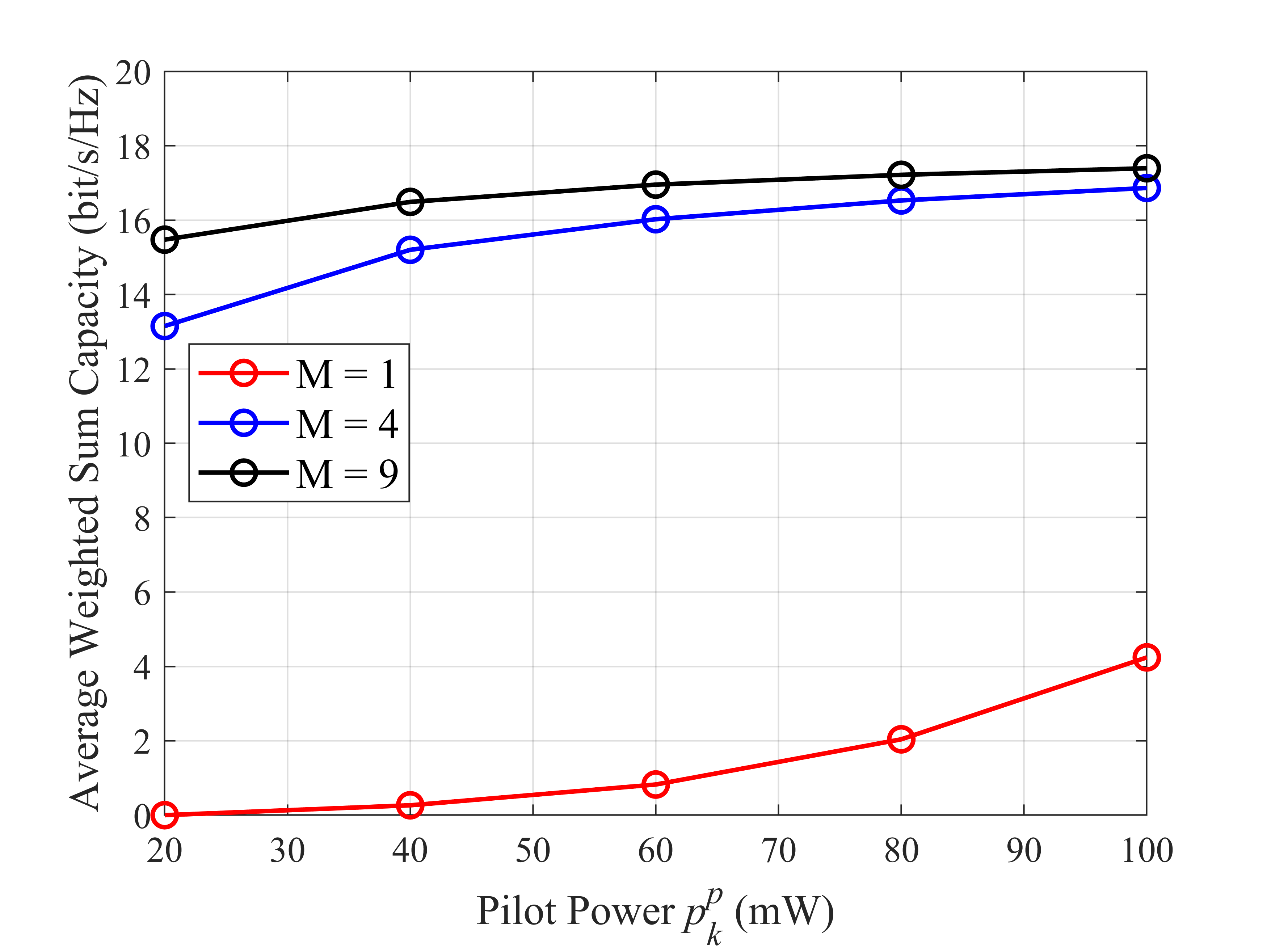}%
\caption*{(a) MRT}
\label{fig:MRC_Pilot_Power}}
\end{minipage}
\hfill
\begin{minipage}[t]{0.32\linewidth}
{\includegraphics[width=2.25in]{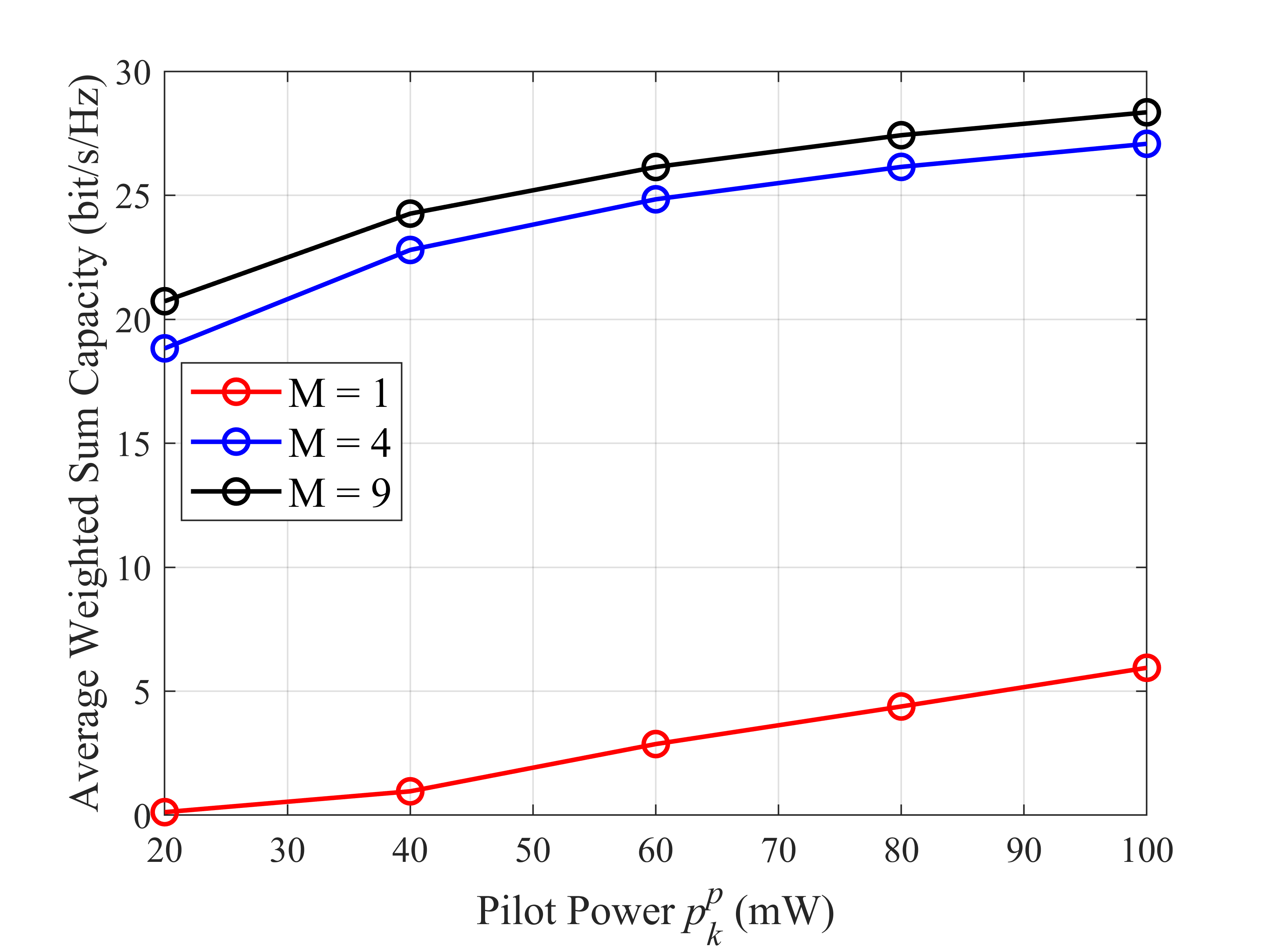}%
\label{fig:FZF_Pilot_Power}}
\caption*{(b) FZF}
\end{minipage}
\hfill
\begin{minipage}[t]{0.32\linewidth}
{\includegraphics[width=2.25in]{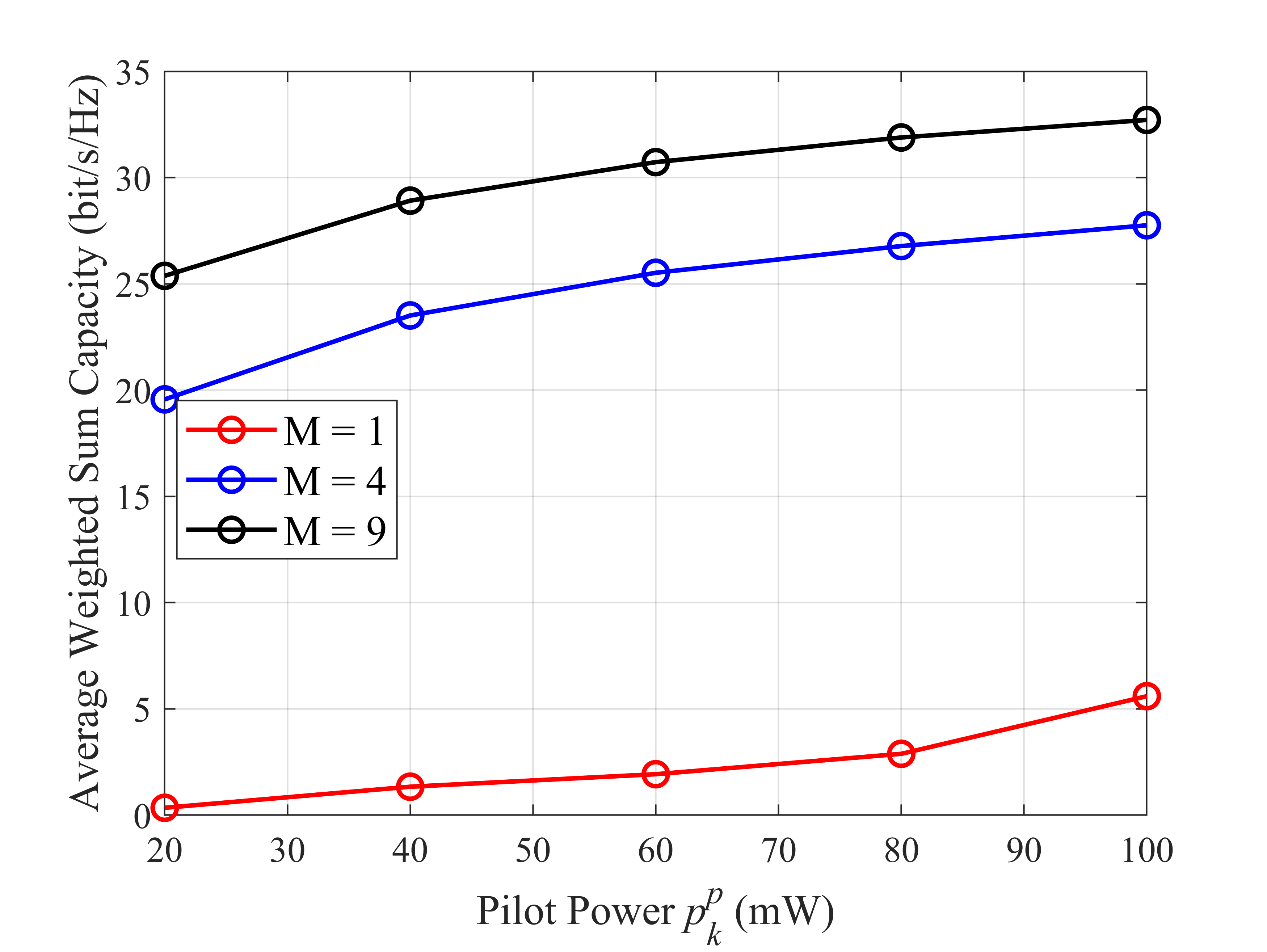}%
\label{fig:LZF_Pilot_Power}}
\caption*{(c) LZF}
\end{minipage}
\caption{Performance of the algorithm versus pilot power for different precoding schemes.}
\label{fig_Pilot_Power}
\end{figure*}

\subsection{Effect of The Number of APs}

\begin{figure}[t]
\centering
\includegraphics[width=2.25in]{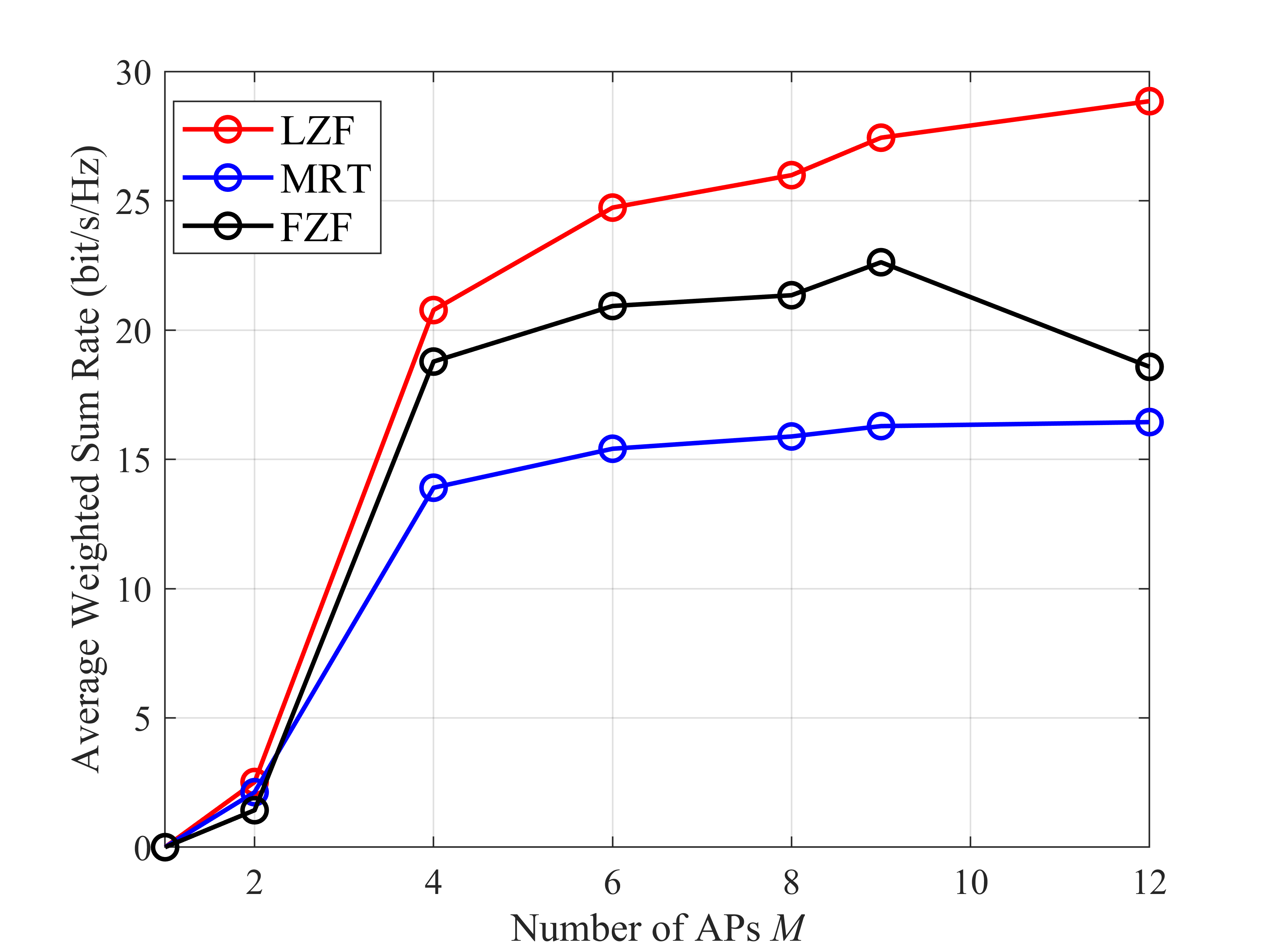}
\caption{Performance of proposed algorithm V.S. Number of APs.}
\label{fig:Vs_M}
\end{figure}
To fully explore the deployment of APs so as to maximize the system performance with limited antennas, we evaluate the average weighted sum rate versus various numbers of APs with $MN = 144$ and $P_m = 0.2$ $W$, $\forall m$ in Fig. \ref{fig:Vs_M}. For the MRT scheme, the average weighted sum rate initially increases with the number of APs, and then it tends to be stable at $16$ $\rm {bit/s/Hz}$. \textcolor{black}{This is due to the fact that each device relying on the MRT precoding scheme becomes interference limited and tends to be stable.} However, for the FZF precoding scheme, the system performance will decrease when the number of APs is large, as deploying more APs causes the reduction in degrees of freedom. In contrast, the system performance using the LZF precoding scheme increases with the number of APs. This is because the LZF precoding scheme strikes a balance between interference suppression and available degrees of freedom, thereby supporting more devices.

\subsection{Effect of The Number of Devices}
\begin{figure*}[t]
\begin{minipage}[t]{0.32\linewidth}
{\includegraphics[width=2.25in]{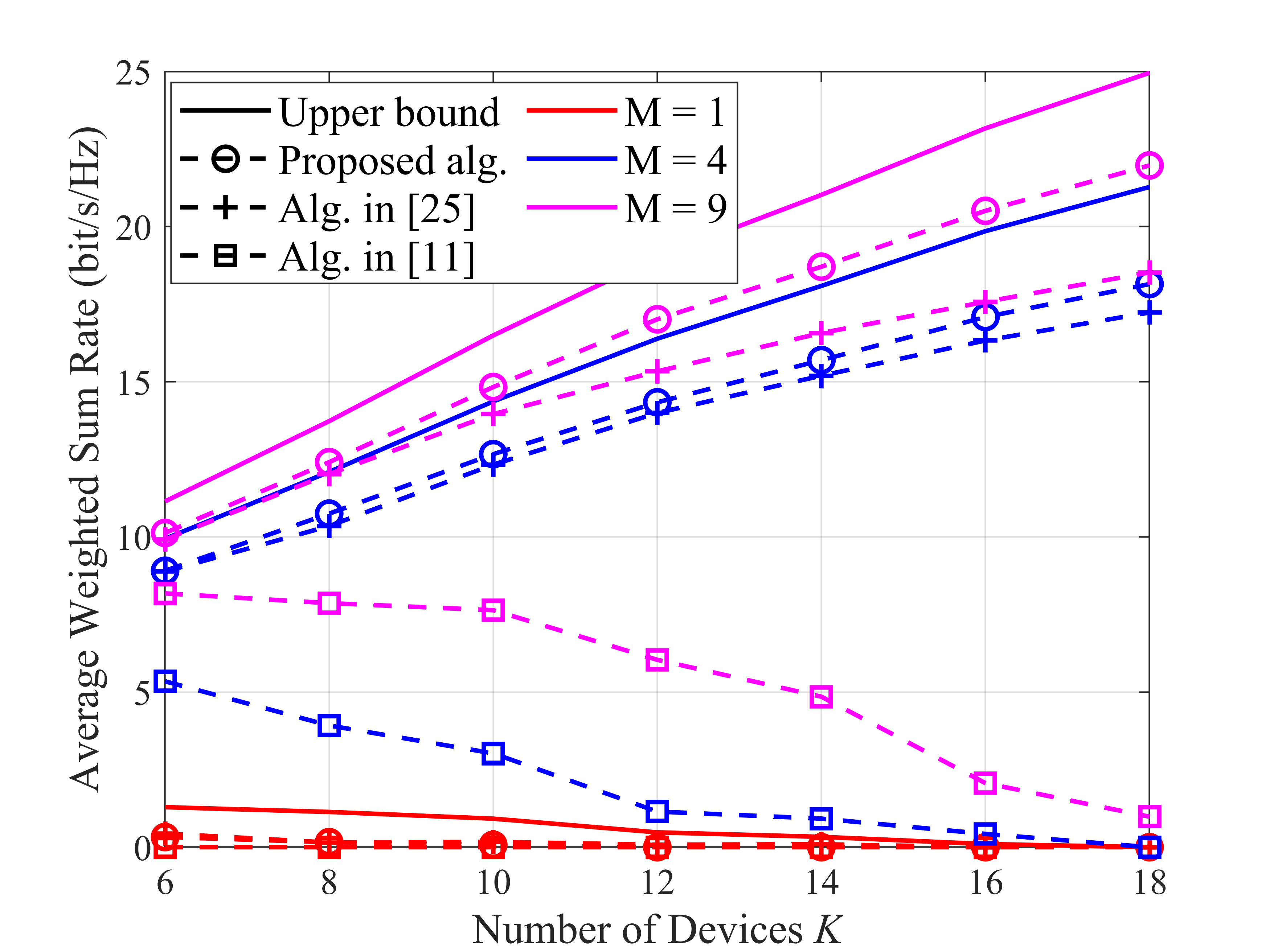}%
\caption*{(a) MRT}
\label{fig:MRC_Vs_Devices}}
\end{minipage}
\hfill
\begin{minipage}[t]{0.32\linewidth}
{\includegraphics[width=2.25in]{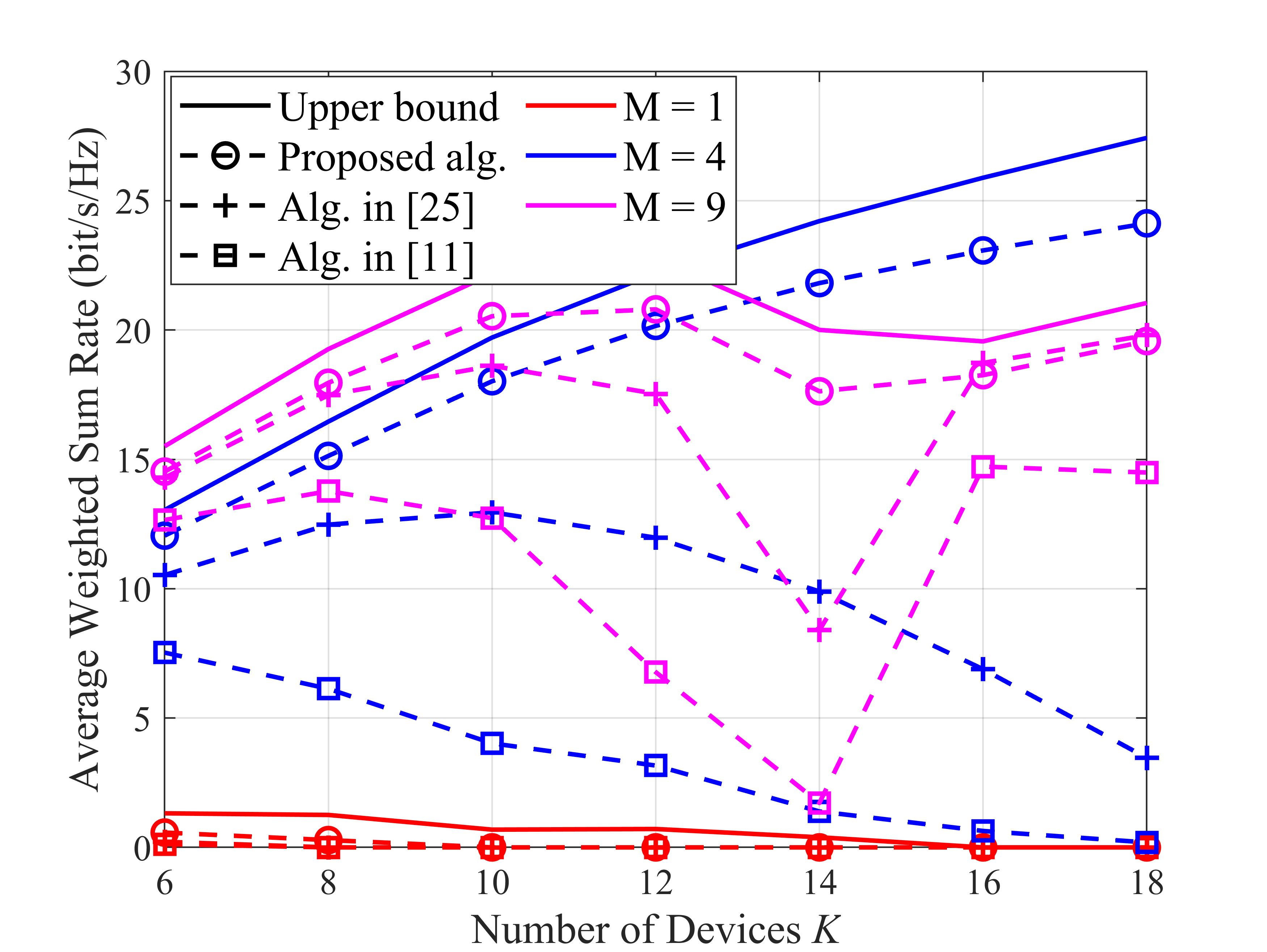}%
\label{fig:FZF_Vs_Devices}}
\caption*{(b) FZF}
\end{minipage}
\hfill
\begin{minipage}[t]{0.32\linewidth}
{\includegraphics[width=2.25in]{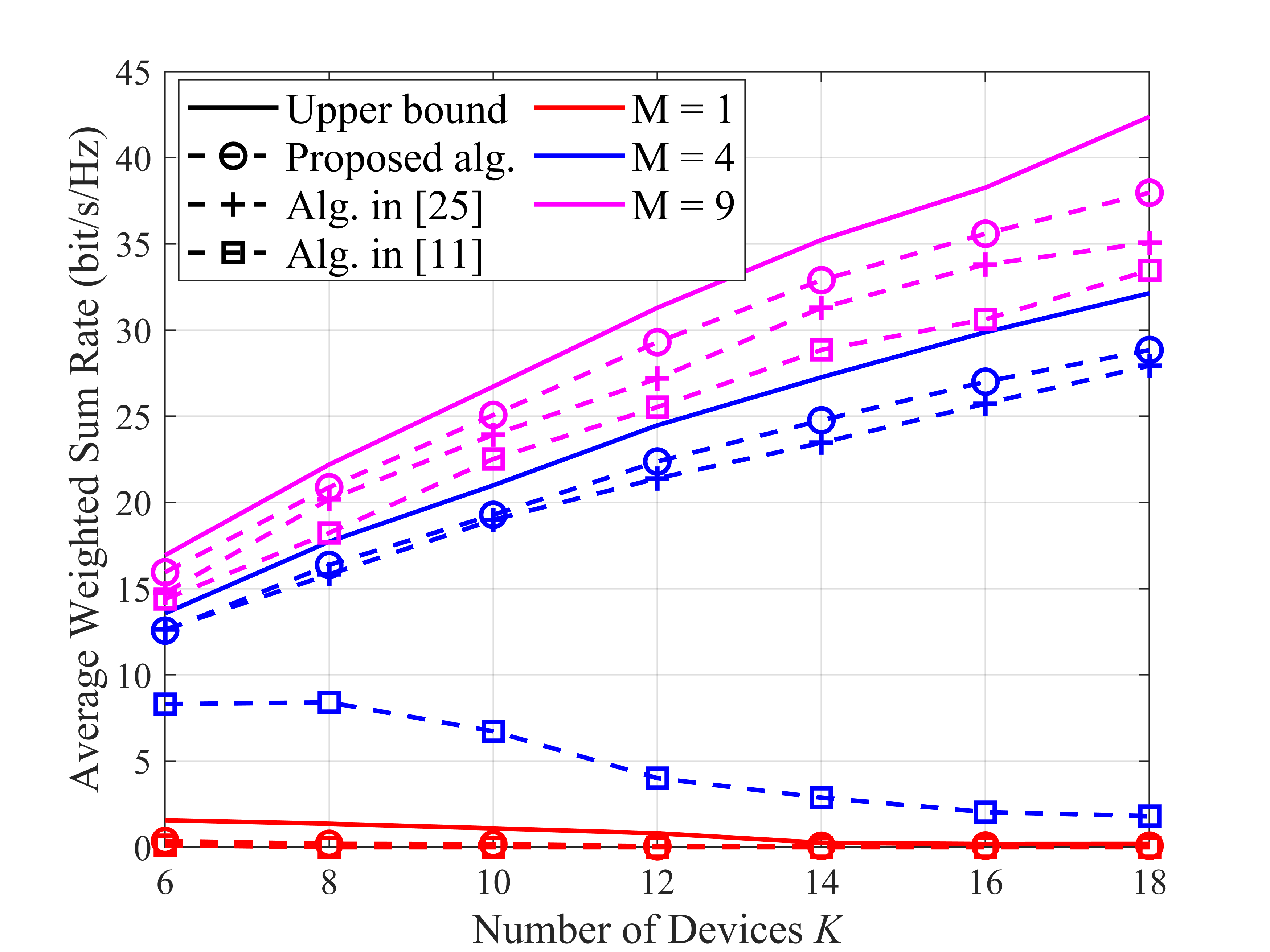}%
\label{fig:LZF_Vs_Devices}}
\caption*{(c) LZF}
\end{minipage}
\caption{\textcolor{black}{Performance of proposed algorithm V.S. Number of devices for different precoding schemes.}}
\label{fig_Devices}
\end{figure*}
\textcolor{black}{To support more devices, a round robin-based scheduler is adopted for the case of $K \ge N$ or $K \ge |{\cal U}_m|$. Specifically, the APs would first transmit signals to the $K_1$ devices, and then serve the remaining $(K - K_1)$ devices in the next time interval. By averaging over 100 random generations, we investigate the relationship between the number of devices and the system performance with $MN = 144$, $K_1 = \frac{K}{2}$, and $P_m = 0.2$ $W$, $\forall m$. To show the effectiveness of our proposed method, the results of the Shannon capacity, the algorithm in \cite{ghanem2019resource}, and power allocation in \cite{ref18} are presented. Furthermore, if any devices violate the requirements, the data rate is set to zero. Obviously, the Shannon capacity is the ideal performance, and the benchmark one in \cite{ghanem2019resource} has an unpredictable trend because it does not consider the penalty due to short packet transmissions. The performance relying on the power allocation of \cite{ref18} can approach that of the proposed method, owing to enhanced path gain. In contrast, the proposed algorithm can approach the upper bound in CF mMIMO systems, which demonstrates the effectiveness of our algorithm. More importantly, the weighted sum rate in the centralized mMIMO is almost zero owing to the failure to meet the requirements, while there is a significant performance improvement in CF mMIMO systems. This is due to the fact that the centralized mMIMO can only support those devices that are close to the APs, instead of all devices, leading to zero data rate. Furthermore, the average weighted sum rate of $M = 9$ APs relying on the FZF precoding increases when $K \le 10$ and then declines when $10 \le K \le 14$. Thereafter, APs based on the FZF precoding scheme can support extra devices with enhanced URLLC services by implementing the scheduler, which motivates us to enforce the appropriate scheduler when the number of served devices approaches that of equipped antennas per AP.}

\section{Conclusion}
In this paper, the resource allocation for a CF mMIMO-enabled URLLC dowlink system was treated. We first derived the closed-form LB data rates with imperfect CSI based on MRT, FZF, and LZF precoding, and maximized the weighted sum rate based on the derived LB data rate. Then, by deriving the globally optimal pilot power and using SCA, the non-convex problem was transformed into a series of subproblems, which can be solved in an iterative manner by our proposed algorithm. Simulation results demonstrated the rapid convergence speed of our algorithm and the optimal AP selection strategy based on \textcolor{black}{the} short packet transmission. \textcolor{black}{Furthermore, the quality of URLLC services will benefit by deploying more APs, except for the FZF precoding scheme. More importantly, the power allocation strategies under the short packet regime can significantly enhance the system performance over the existing algorithms.}

\textcolor{black}{Regarding CF mMIMO systems, it is impractical to assign orthogonal pilot sequences to multiple devices under the FCBL. Therefore, investigating the pilot allocation scheme and analyzing the impact of sharing pilot sequences would be left for our future work. Furthermore, since it is unrealistic to assume an idealized fronthaul link between the CPU and the APs, the limited fronthaul will be studied in the future.}

\bibColoredItems{black}{ghanem2019resource,ref18,zhang2020prospective,zhang2021improving,wang2022uplink,zhang2018performance,luo2022downlink,van2018joint,mehanna2014feasible,ref16}

\begin{appendices}
	\section{Proof of Theorem \ref{MRC_SINR_T}}
	\label{MRC_SINR_P}
	Before proving this theorem, we need to calculate the precoding vector for the MRT case. The normalized precoding vector is given by
	\begin{equation}
		\setlength\abovedisplayskip{5pt}
		\setlength\belowdisplayskip{5pt}
		\label{MRC_normalized}
		\begin{split}
			{\bf{a}}_{m,k}^{{\rm{MRT}}} &= \frac{{{\alpha _{m,k}}\left( {{{\bf{g}}_{m,k}} + {\bf{n}}_{m,k}^p} \right)}}{{\sqrt {{{\mathbb{E}\left\| {{\alpha _{m,k}}\left( {{{\bf{g}}_{m,k}} + {\bf{n}}_{m,k}^p} \right)} \right\|}^2}} }} = \frac{{{{\bf{g}}_{m,k}} + {\bf{n}}_{m,k}^p}}{{\sqrt {N\left( {{\beta _{m,k}} + \frac{1}{{Kp_k^p}}} \right)} }} \\
			& = \frac{{\sqrt {{\lambda _{m,k}}} }}{{{\beta _{m,k}}\sqrt N }}\left( {{{\bf{g}}_{m,k}} + {\bf{n}}_{m,k}^p} \right),
		\end{split}
	\end{equation}
	where $\alpha _{m,k}$ is $\alpha _{m,k} = \frac{Kp_k^p \beta_{m,k}}{Kp_k^p \beta_{m,k} + 1}$.
	
	Then, we need to derive the expressions of ${\left| {{\rm{DS}}_{k}} \right|^2}$, $\mathbb{E}\left( \left| {{\rm{LS}}_{k}} \right|^2 \right)$, $\mathbb{E}\left( \left| {{\rm{UI}}_{k,k'}} \right|^2 \right)$ and $\mathbb{E}\left( \left| {{\rm{N}}_{k}} \right|^2 \right)$, respectively. We first compute ${\rm{DS}}_k$. Since $\hat {\bf{g}}_{m,k}$ and $\tilde {\bf{g}}_{m,k}$ are independent, we have
	\begin{equation}
		\setlength\abovedisplayskip{5pt}
		\setlength\belowdisplayskip{5pt}
		\label{MRC_DSk}
		\begin{split}
			&{\left| {{\rm{DS}}_{k}} \right|^2} \\
			&= {\left|\mathbb{E} {\left\{ {\sum\limits_{m \in {{\cal M}_k}} {\sqrt {p_{m,k}^d} {{\left( {{{\bf{g}}_{m,k}}} \right)}^T}{{\left( {{\bf{a}}_{m,k}^{{\rm{MRT}}}} \right)}^ * }} } \right\}} \right|^2} \\
			& = {\left|\mathbb{E} {\left\{ {\sum\limits_{m \in {{\cal M}_k}} {\sqrt {p_{m,k}^d} {{\left( {{{\bf{g}}_{m,k}}} \right)}^T}{{ {\frac{{\sqrt {{\lambda _{m,k}}} }}{{{\beta _{m,k}}\sqrt N }}\left( {{{\bf{g}}_{m,k}} + {\bf{n}}_{m,k}^p} \right)}}^ * }} } \right\}} \right|^2} \\
			& = {\left| {\sum\limits_{m \in {{\cal M}_k}} {\sqrt {Np_{m,k}^d {\lambda _{m,k}}} } } \right|^2}.
		\end{split}
	\end{equation}

	The term $\mathbb{E}\left( \left| {{\rm{LS}}_{k}} \right|^2 \right)$ is given by
	\begin{equation}
		\setlength\abovedisplayskip{5pt}
		\setlength\belowdisplayskip{5pt}
		\label{MRC_LSk}
		\begin{split}
			& \mathbb{E} \left\{ {{{\left| {{\rm{LS}}_{k}} \right|}^2}} \right\} \\  
			& = \mathbb{E} \left\{{{{\left| {\sum\limits_{m \in {{\cal M}_k}} {\sqrt {p_{m,k}^d} {{\left( {{{\bf{g}}_{m,k}}} \right)}^T}{{\left( {{\bf{a}}_{m,k}^{{\rm{MRT}}}} \right)}^ * }}  - \rm{DS}_k} \right|}^2}} \right\} \\
			&+ \mathbb{E} \left\{ {{\left| {\sum\limits_{m \in {{\cal M}_k}} {\frac{{\sqrt {p_{mk}^d{\lambda _{m,k}}} }}{{{\beta _{m,k}}\sqrt N }}{{\left( {{{\bf{g}}_{m,k}}} \right)}^T}{{\left( {{\bf{n}}_{m,k}^p} \right)}^*}} } \right|}^2}\right\} \\
			&- {\left( {\sum\limits_{m \in {{\cal M}_k}} {\sqrt {N{\lambda _{m,k}}p_{m,k}^d} } } \right)^2} \\
			& = \sum\limits_{m \in {{\cal M}_k}} {p_{m,k}^d{\beta _{m,k}}}.
		\end{split}
	\end{equation}
	
	Then, $\mathbb{E}\left( \left| {{\rm{UI}}_{k,k'}} \right|^2 \right)$ can be calculated as
	\begin{equation}
		\setlength\abovedisplayskip{5pt}
		\setlength\belowdisplayskip{5pt}
		\label{MRC_UIkk}
		\begin{split}
			& \mathbb{E} \left( {{{\left| {{\rm{U}}{{\rm{I}}_{k,k'}}} \right|}^2}} \right) \\
			& = \mathbb{E} \left\{ {{{\left| {\sum\limits_{m \in {{\cal M}_{k'}}} {\frac{{\sqrt {p_{m,k'}^d{\gamma _{m,k'}}} }}{{{\beta _{m,k'}}\sqrt N }}{{\left( {{{\bf{g}}_{m,k}}} \right)}^T}{{\left( {{{\bf{g}}_{m,k'}}} \right)}^ * }} } \right|}^2}} \right\}  \\
			& +\mathbb{E} \left\{{{\left| {\sum\limits_{m \in {{\cal M}_{k'}}} {\frac{{\sqrt {p_{m,k'}^d{\gamma _{m,k'}}} }}{{{\beta _{m,k'}}\sqrt N }}{{\left( {{{\bf{g}}_{m,k}}} \right)}^T}{{\left( {{\bf{n}}_{m,k'}^p} \right)}^ * }} } \right|}^2} \right\}.
		\end{split}
	\end{equation}
	For each term in (\ref{MRC_UIkk}), we have
	\begin{equation}
		\setlength\abovedisplayskip{5pt}
		\setlength\belowdisplayskip{5pt}
		\label{UIkk_first}
		\begin{split}
			&\mathbb{E} \left\{ {{{\left| {\sum\limits_{m \in {{\cal M}_{k'}}} {\frac{{\sqrt {p_{m,k'}^d{\gamma _{m,k'}}} }}{{{\beta _{m,k'}}\sqrt N }}{{\left( {{{\bf{g}}_{m,k}}} \right)}^T}{{\left( {{{\bf{g}}_{m,k'}}} \right)}^ * }} } \right|}^2}} \right\} \\
			& = \sum\limits_{m \in {{\cal M}_{k'}}} {p_{m,k'}^d{\gamma _{m,k'}}} \frac{{{\beta _{m,k}}}}{{{\beta _{m,k'}}}}
		\end{split}
	\end{equation}
	and
	\begin{equation}
		\setlength\abovedisplayskip{5pt}
		\setlength\belowdisplayskip{5pt}
		\label{UIkk_second}
		\begin{split}
			&\mathbb{E} \left\{ {{{\left| {\sum\limits_{m \in {{\cal M}_{k'}}} {\frac{{\sqrt {p_{m,k'}^d{\gamma _{m,k'}}} }}{{{\beta _{m,k'}}\sqrt N }}{{\left( {{{\bf{g}}_{m,k}}} \right)}^T}{{\left( {{\bf{n}}_{m,k'}^p} \right)}^ * }} } \right|}^2}} \right\} \\
			& = \sum\limits_{m \in {{\cal M}_{k'}}} {p_{m,k'}^d{\gamma _{m,k'}}\frac{1}{{Kp_{_{k'}}^p}}} \frac{{{\beta _{mk}}}}{{{{\left( {{\beta _{mk'}}} \right)}^2}}} .
		\end{split}
	\end{equation}
	By combining (\ref{UIkk_first}) with (\ref{UIkk_second}), we have
	\begin{align}
		\setlength\abovedisplayskip{5pt}
		\setlength\belowdisplayskip{5pt}
		\label{MRC_UIkk_final}
		\setlength\abovedisplayskip{5pt}
		\setlength\belowdisplayskip{5pt}
		& \mathbb{E}\left( \left| {{\rm{UI}}_{k,k'}} \right|^2 \right) \notag \\
		&= \sum\limits_{m \in {{\cal M}_{k'}}} {p_{m,k'}^d{\gamma _{m,k'}}} \frac{{{\beta _{m,k}}}}{{{\beta _{m,k'}}}} + \sum\limits_{m \in {{\cal M}_{k'}}} {p_{m,k'}^d{\gamma _{m,k'}}\frac{{{\beta _{m,k}}}}{{{{\left( {{\beta _{m,k'}}} \right)}^2}}}\frac{1}{{Kp_{_{k'}}^p}}} \notag \\
		& = \sum\limits_{m \in {{\cal M}_{k'}}} {p_{m,k'}^d{\beta _{m,k}}}.
	\end{align}
	
	Finally, we compute $\mathbb{E}\left( \left| {{\rm{N}}_{k}} \right|^2 \right)$, which is written as
	\begin{equation}
		\setlength\abovedisplayskip{5pt}
		\setlength\belowdisplayskip{5pt}
		\label{MRC_Nk}
		\mathbb{E} \left\{ {{{\left| {{n_k}} \right|}^2}} \right\} = 1.
	\end{equation}
	
	Substituting (\ref{MRC_DSk}), (\ref{MRC_LSk}), (\ref{MRC_UIkk_final}), and (\ref{MRC_Nk}) into (\ref{kth_SINR_downlink}), we obtain $\hat \gamma _k^{{\rm{MRT}}}$ in (\ref{MRC_SINR_LB}).
	
	\section{Proof of Theorem \ref{FZF_SINR_T}}
	\label{FZF_SINR_P}
	Before proving this theorem, we need to provide the precoding vector. By using the identity \cite{2004Random}, the normalized coefficient can be derived as
	\begin{align}
		\setlength\abovedisplayskip{5pt}
		\setlength\belowdisplayskip{5pt}
		\label{Normalized_ZF}
		& {{{\mathbb{E}} \left\{ {{{\left\| {{{{\bf{\hat G}}}_m}{{\left[ {{\bf{\hat G}}_m^H{{{\bf{\hat G}}}_m}} \right]}^{ - 1}}{{\bf{e}}_k}} \right\|}^2}} \right\}}} \notag\\
		&= {\mathbb{E}} \left\{ {{{\left( {{{\bf{e}}_k}} \right)}^H}{{\left[ {{\bf{\hat G}}_m^H{{{\bf{\hat G}}}_m}} \right]}^{ - 1}}{{\left( {{{{\bf{\hat G}}}_m}} \right)}^H}{{{\bf{\hat G}}}_m}{{\left[ {{\bf{\hat G}}_m^H{{{\bf{\hat G}}}_m}} \right]}^{ - 1}}{{\bf{e}}_k}} \right\} \notag \\
		& = {\mathbb{E}}  \left\{ {{{\left( {{{\bf{e}}_k}} \right)}^H}{{\left[ {{\bf{\hat G}}_m^H{{{\bf{\hat G}}}_m}} \right]}^{ - 1}}{{\bf{e}}_k}} \right\} \notag \\
		& =  \frac{1}{{\left( {N - K} \right){\lambda _{m,k}}}}.
	\end{align}
	
	Then, ${\left| {{\rm{DS}}_{k}} \right|}$ can be derived as
	\begin{equation}
		\setlength\abovedisplayskip{5pt}
		\setlength\belowdisplayskip{5pt}
		\label{DS_ZF}
		\begin{split}
			& \left| {{\rm{DS}}_{k}} \right|^2 \\
			&= {\left| \mathbb{E}{\left\{ {\sum\limits_{m \in {{\cal M}_k}} {{{\left( {{{\bf{g}}_{m,k}}} \right)}^T}{{\left( {{\bf{a}}_{m,k}^{{\rm{FZF}}}} \right)}^ * }\sqrt {p_{m,k}^d} } } \right\}} \right|^2} \\
			& = {\left| \mathbb{E}{\left\{ {\sum\limits_{m \in {{\cal M}_k}} {{{\left( {{{{\bf{\hat g}}}_{m,k}} + {{{\bf{\tilde g}}}_{m,k}}} \right)}^T}{{\left( {{\bf{a}}_{m,k}^{{\rm{FZF}}}} \right)}^ * }\sqrt {p_{m,k}^d} } } \right\}} \right|^2} \\
			& = {\left( {\sum\limits_{m \in {{\cal M}_k}} {\sqrt {\left( {N - K} \right)p_{m,k}^d{\lambda _{m,k}}} } } \right)^2}.
		\end{split}
	\end{equation}

	Next, the leakage power can be formulated as
	\begin{equation}
		\setlength\abovedisplayskip{5pt}
		\setlength\belowdisplayskip{5pt}
		\label{LS_ZF}
		\begin{split}
			& \mathbb{E} \left\{ {{{\left| {{\rm{LS}}_{k}} \right|}^2}} \right\} \\
			& = \mathbb{E}\left\{ {{{\left| {\sum\limits_{m \in {{\cal M}_k}} {{{\left( {{{\bf{g}}_{m,k}}} \right)}^T}{{\left( {{\bf{a}}_{m,k}^{{\rm{FZF}}}} \right)}^ * }\sqrt {p_{m,k}^d} }  - \rm{DS}_k} \right|}^2}} \right\}\\
			& = \mathbb{E} \left\{ {{{\left| {\sum\limits_{m \in {{\cal M}_k}} {{{\left( {{{{\bf{\tilde g}}}_{m,k}}} \right)}^T}{{\left( {{\bf{a}}_{m,k}^{{\rm{FZF}}}} \right)}^*}\sqrt {p_{m,k}^d} } } \right|}^2}} \right\} \\
			& = \sum\limits_{m \in {{\cal M}_k}} {p_{m,k}^d\left( {{\beta _{m,k}} - {\lambda _{m,k}}} \right)}.
		\end{split}
	\end{equation}
	
	The term $\mathbb{E}\left( \left| {{\rm{UI}}_{k,k'}} \right|^2 \right)$ can be expressed as
	\begin{equation}
		\setlength\abovedisplayskip{5pt}
		\setlength\belowdisplayskip{5pt}
		\label{UI_ZF}
		\begin{split}
			& \mathbb{E} \left\{ {{{\left| {{\rm{UI}}_{k,k'}} \right|}^2}} \right\} \\
			&= \mathbb{E} \left\{ {{{\left| {\sum\limits_{m \in {{\cal M}_{k'}}} {{{\left( {{{\bf{g}}_{m,k}}} \right)}^T}{{\left( {{\bf{a}}_{m,k'}^{{\rm{FZF}}}} \right)}^ * }\sqrt {p_{m,k'}^d} } } \right|}^2}} \right\} \\
			& = \mathbb{E} \left\{ {{{\left| {\sum\limits_{m \in {{\cal M}_{k'}}} {{{\left( {{{{\bf{\tilde g}}}_{m,k}}} \right)}^T}{{\left( {{\bf{a}}_{m,k'}^{{\rm{FZF}}}} \right)}^ * }\sqrt {p_{m,k'}^d} } } \right|}^2}} \right\} \\
			& = \sum\limits_{m \in {{\cal M}_{k'}}} {p_{m,k'}^d\left( {{\beta _{m,k}} - {\lambda _{m,k}}} \right)}.
		\end{split}
	\end{equation}
	
	Finally, we complete the proof by substituting the expressions of (\ref{DS_ZF}), (\ref{LS_ZF}), (\ref{UI_ZF}), and $ \mathbb{E} \left\{ {{{\left| {{n_k}} \right|}^2}} \right\} = 1$ into the SINR expression.
	
	\section{Proof of Theorem \ref{LZF_SINR_T}}
	\label{LZF_SINR_P}
	The normalized coefficient can be derived as
	\begin{align}
		\setlength\abovedisplayskip{5pt}
		\setlength\belowdisplayskip{5pt}
		\label{Normalized_LZF}
		& \mathbb{E} \left\{ {{{\left\| {{{{\bf{\hat G}}}_m}{{\bf{E}}_{{{\cal U}_m}}}{{\left( {{\bf{E}}_{{{\cal U}_m}}^H{\bf{\hat G}}_m^H{{{\bf{\hat G}}}_m}{{\bf{E}}_{{{\cal U}_m}}}} \right)}^{ - 1}}{{\bm{\xi }}_{m,k}}} \right\|}^2}} \right\} \notag \\
		& = \mathbb{E} \left\{ {{{\left( {{{\bm{\xi }}_{m,k}}} \right)}^H}{{\left( {{\bf{E}}_{{{\cal U}_m}}^H{\bf{\hat G}}_m^H{{{\bf{\hat G}}}_m}{{\bf{E}}_{{{\cal U}_m}}}} \right)}^{ - 1}}{{\bm{\xi }}_{m,k}}} \right\} \notag \\
		& = \frac{1}{{\left( {N - {\tau _m}} \right){\lambda _{m,k}}}},
	\end{align}
	where $\tau_m$ is defined in (\ref{t_m}).
	
	Then, the desired signal ${\left| {{\rm{DS}}_{k}} \right|^2}$ can be given by
		\begin{equation}
			\setlength\abovedisplayskip{5pt}
			\setlength\belowdisplayskip{5pt}
			\label{DS_LZF}
			\begin{split}
				& \left| {{\rm{DS}}_{k}} \right|^2 \\
				&= \left|\mathbb{E} \left\{ {\sum\limits_{m \in {{\cal M}_k}} {{{\left( {{{\bf{g}}_{m,k}}} \right)}^T}{{\left( {{\bf{a}}_{m,k}^{{\rm{LZF}}}} \right)}^ * }\sqrt {p_{m,k}^d} } } \right\}\right|^2 \\
				& = \left|\mathbb{E}\left\{ {\sum\limits_{m \in {{\cal M}_k}} {\sqrt {\left( {N - {\tau _m}} \right){\lambda _{m,k}}p_{m,k}^d} {{\left( {{{{\bf{\hat g}}}_{m,k}}} \right)}^T}{{\left( {{{{\bf{\hat G}}}_m}{{\bf{E}}_{{{\cal U}_m}}}{{\left( {{\bf{E}}_{{{\cal U}_m}}^H{\bf{\hat G}}_m^H{{{\bf{\hat G}}}_m}{{\bf{E}}_{{{\cal U}_m}}}} \right)}^{ - 1}}{{\bm{\xi }}_{m,k}}} \right)}^ * }} } \right\}\right|^2 \\
				&= \left(\sum\limits_{m \in {{\cal M}_k}} {\sqrt {\left( {N - {\tau _m}} \right)p_{m,k}^d{\lambda _{m,k}}} }\right)^2.
			\end{split}
		\end{equation}

	Next, similar to the FZF case, the leakage power for the AP using the LZF precoding scheme can be formulated as
	\begin{equation}
		\setlength\abovedisplayskip{5pt}
		\setlength\belowdisplayskip{5pt}
		\label{LS_LZF}
		\begin{split}
			&\mathbb{E} \left\{ {{{\left| {{\rm{LS}}_{k}} \right|}^2}} \right\} \\
			&= \mathbb{E}\left\{ {{{\left| {\sum\limits_{m \in {{\cal M}_k}} {{{\left( {{{\bf{g}}_{m,k}}} \right)}^T}{{\left( {{\bf{a}}_{m,k}^{{\rm{LZF}}}} \right)}^ * }\sqrt {p_{m,k}^d} }  - \rm{DS}_k} \right|}^2}} \right\}\\
			& = \mathbb{E} \left\{ {{{\left| {\sum\limits_{m \in {\mathcal{M}_k}} {{{\left( {{{{\bf{\tilde g}}}_{m,k}}} \right)}^T}{{\left( {{\bf{a}}_{m,k}^{{\rm{LZF}}}} \right)}^*}\sqrt {p_{m,k}^d} } } \right|}^2}} \right\}\\
			& = \sum\limits_{m \in {{\cal M}_k}} {p_{m,k}^d\left( {{\beta _{m,k}} - {\lambda _{m,k}}} \right)}.
		\end{split}
	\end{equation}
	
	The term of the devices' interference is different from that of the FZF scheme, as the interference from other devices may not be suppressed. The term $\mathbb{E}\left( \left| {{\rm{UI}}_{k,k'}} \right|^2 \right)$ can be given by 
		\begin{equation}
			\setlength\abovedisplayskip{5pt}
			\setlength\belowdisplayskip{5pt}
			\label{UI_LZF}
			\begin{split}
				& \mathbb{E} \left\{ {{{\left| {{\rm{UI}}_{k,k'}} \right|}^2}} \right\} = \mathbb{E} \left\{ {{{\left| {\sum\limits_{m \in {{\cal M}_{k'}}} {{{\left( {{{\bf{g}}_{m,k}}} \right)}^T}{{\left( {{\bf{a}}_{m,k'}^{{\rm{LZF}}}} \right)}^ * }\sqrt {p_{m,k'}^d} } } \right|}^2}} \right\} \\
				& = \mathbb{E} \left\{ {{{\left| {\sum\limits_{m \in \left\{ {{{\cal M}_k} \cap {{\cal M}_{k'}}} \right\}} {{{\left( {{{\bf{g}}_{m,k}}} \right)}^T}{{\left( {{\bf{a}}_{m,k'}^{{\rm{LZF}}}} \right)}^ * }\sqrt {p_{m,k'}^d} }  + \sum\limits_{m \in \left\{ {{{\cal M}_{k'}}\backslash \left\{ {{{\cal M}_k} \cap {{\cal M}_{k'}}} \right\}} \right\}} {{{\left( {{{\bf{g}}_{m,k}}} \right)}^T}{{\left( {{\bf{a}}_{m,k'}^{{\rm{LZF}}}} \right)}^ * }\sqrt {p_{m,k'}^d} } } \right|}^2}} \right\} \\
				& = \mathbb{E} \left\{ {{{\left| {\sum\limits_{m \in \left\{ {{{\cal M}_k} \cap {{\cal M}_{k'}}} \right\}} {{{\left( {{{\bf{g}}_{m,k}}} \right)}^T}{{\left( {{\bf{a}}_{m,k'}^{{\rm{LZF}}}} \right)}^ * }\sqrt {p_{m,k'}^d} } } \right|}^2} + {{\left| {\sum\limits_{m \in \left\{ {{{\cal M}_{k'}}\backslash \left\{ {{{\cal M}_k} \cap {{\cal M}_{k'}}} \right\}} \right\}} {{{\left( {{{\bf{g}}_{m,k}}} \right)}^T}{{\left( {{\bf{a}}_{m,k'}^{{\rm{LZF}}}} \right)}^ * }\sqrt {p_{m,k'}^d} } } \right|}^2}} \right\}.
			\end{split}
		\end{equation}

	As can be seen from (\ref{UI_LZF}), the devices' interference consists of two terms. In specific, the first term means that the vector $\hat {\bf g}_{m,k}$ is chosen by the selection matrix ${\bf{E}}_{{{\cal U}_m}}$ and the second term means that not chosen by matrix ${\bf{E}}_{{{\cal U}_m}}$. Obviously, the interference of the first term can be suppressed as $\hat {\bf g}_{m,k} {{\bf{a}}_{m,k'}^{{\rm{LZF}}}}$ is equal to zero, while the second term's interference cannot be suppressed.
	
	Then, the first and the second terms of $\mathbb{E}\left( \left| {{\rm{UI}}_{k,k'}} \right|^2 \right)$ are given by
	\begin{equation}
		\setlength\abovedisplayskip{5pt}
		\setlength\belowdisplayskip{5pt}
		\label{UI_LZF_first}
		\begin{split}
			&\mathbb{E} \left\{ {{{\left| {\sum\limits_{m \in \left\{ {{{\cal M}_k} \cap {{\cal M}_{k'}}} \right\}} {{{\left( {{{\bf{g}}_{m,k}}} \right)}^T}{{\left( {{\bf{a}}_{m,k'}^{{\rm{LZF}}}} \right)}^ * }\sqrt {p_{m,k'}^d} } } \right|}^2}} \right\} \\
			& = \mathbb{E}  \left\{ {{{\left| {\sum\limits_{m \in \left\{ {{{\cal M}_k} \cap {{\cal M}_{k'}}} \right\}} {{{\left( {{{{\bf{\tilde g}}}_{m,k}}} \right)}^T}{{\left( {{\bf{a}}_{m,k'}^{{\rm{LZF}}}} \right)}^ * }\sqrt {p_{m,k'}^d} } } \right|}^2}} \right\} \\
			& = \sum\limits_{m \in \left\{ {{{\cal M}_k} \cap {{\cal M}_{k'}}} \right\}} {p_{m,k'}^d\left( {{\beta _{m,k}} - {\lambda _{m,k}}} \right)},
		\end{split}
	\end{equation}
	and
	\begin{equation}
		\setlength\abovedisplayskip{5pt}
		\setlength\belowdisplayskip{5pt}
		\label{UI_LZF_second}
		\begin{split}
			&\mathbb{E} \left\{ {{{\left| {\sum\limits_{m \in \left\{ {{{\cal M}_{k'}}\backslash \left\{ {{{\cal M}_k} \cap {{\cal M}_{k'}}} \right\}} \right\}} {{{\left( {{{\bf{g}}_{m,k}}} \right)}^T}{{\left( {{\bf{a}}_{m,k'}^{{\rm{LZF}}}} \right)}^ * }\sqrt {p_{m,k'}^d} } } \right|}^2}} \right\} \\
			&= \sum\limits_{m \in \left\{ {{{\cal M}_{k'}}\backslash \left\{ {{{\cal M}_k} \cap {{\cal M}_{k'}}} \right\}} \right\}} {p_{m,k'}^d{\beta _{m,k}}}.
		\end{split}
	\end{equation}

	Finally, the $k$th device's SINR using the LZF precoding scheme is obtained by substituting the expressions of (\ref{DS_LZF}), (\ref{LS_LZF}), (\ref{UI_LZF}), and $ \mathbb{E} \left\{ {{{\left| {{n_k}} \right|}^2}} \right\} = 1$ into (\ref{kth_SINR_downlink}).
	
	\section{Proof of Lemma \ref{maximal_pp}}
	\label{proof_pp}
	We first derive the first-order derivative of $f_k(\frac{1}{\hat \gamma_k})$, denoted as
	\begin{equation}
		\setlength\abovedisplayskip{5pt}
		\setlength\belowdisplayskip{5pt}
		\label{Y_derivative}
		\frac{{d{f_k}\left( {\frac{1}{{{{\hat \gamma }_k}}}} \right)}}{{dp_k^p}} = \frac{{ - {f_k}^\prime \left( {\frac{1}{{{{\hat \gamma }_k}}}} \right)}}{{{{\left( {{{\hat \gamma }_k}} \right)}^2}}}\frac{{d{{\hat \gamma }_k}}}{{dp_k^p}}.
	\end{equation}
	
	As can be seen, the sign of the first-order derivative depends on the sign of $\frac{{d{{\hat \gamma }_k}}}{{dp_k^p}}$. Due to the different SINR expressions of the three precoding schemes, we define a general expression ${\hat \gamma }_k = Y(\lambda_{m,k})$. Then, it is readily to prove that ${\hat \gamma }_k$ monotonically increases with $\lambda_{m,k}$, and the first-order derivative of $\lambda_{m,k}$ is $\frac{{K{{\left( {{\beta _{m,k}}} \right)}^2}}}{{{{\left( {Kp_k^p{\beta _{m,k}} + 1} \right)}^2}}} > 0$. Therefore, the function $f_k(\frac{1}{\hat \gamma_k})$ monotonically increases with pilot power $p_k^p$, and the data rate can be maximized when $p_k^p = P_k^{{\rm max},p}$.

	\section{Proof of Lemma \ref{lnx}}
	\label{proof_lnx}
	The inequality in (\ref{lemma1}) can be readily proved by substituting the expressions of $\rho$ and $\delta$ into (\ref{lemma1}). Then, we define $J\left( x \right) = \ln \left( {1 + x} \right) - \rho \ln x - \delta$, the first-order derivative is given by
	\begin{equation}
		\setlength\abovedisplayskip{5pt}
		\setlength\belowdisplayskip{5pt}
		\label{J_derivative}
		\frac{{dJ\left( x \right)}}{{dx}} = \frac{{x - \rho \left( {1 + x} \right)}}{{\left( {1 + x} \right)x}} = \frac{{x\left( {1 + \hat x} \right) - \hat x \left( {1 + x} \right)}}{{\left( {1 + \hat x} \right)\left( {1 + x} \right)x}}.
	\end{equation}
	
	Since both $x$ and $\hat x$ are positive values, the sign of $\frac{{dJ\left( x \right)}}{{dx}}$ only depends on the numerator. Let us define $H(x) = {x\left( {1 + \hat x} \right) - \hat x \left( {1 + x} \right)}$, and then the first-order derivative of $H(x)$ is given by $H'(x) = 1$, which means $H(x)$ monotonically increases. Consequently, since $H(\hat x)$ is equal to zero, we have $H(x) \ge 0$ when $x \ge \hat x$ and $H(x) \le 0$ when $x \le \hat x$, which indicates that $J\left( x \right)$ is an increasing function when $x \ge \hat x$ and a decreasing function when $x \le \hat x$. As a result, we complete the proof by showing that $J\left( x\right)$ is always larger than $J\left( \hat x\right) = 0$.

	\section{Proof of Theorem \ref{theta_T}}
	\label{proof_theta_T}
	By taking the logarithm operator for the left hand side of (\ref{MRC_theta_LB}), we have
	\begin{equation}
		\setlength\abovedisplayskip{5pt}
		\setlength\belowdisplayskip{5pt}
		\ln \left( {{\theta _k}} \right) = \ln \left( { {\sum\limits_{m \in {{\cal M}_k}} {\sqrt {\left( {N - t_m} \right)p_{m,k}^d{\hat \lambda _{m,k}}} } } }\right) \triangleq F\left( \bf x \right), \label{Fx}
	\end{equation}
	where $ \bf x$ is given by $ {\bf x} = \left[\ln [\left(N - t_1\right)  p_{1,k}^d {\hat \lambda _{1,k}} ],\cdot \cdot \cdot,\ln [\left(N - t_m\right)  p_{m,k}^d {\hat \lambda _{m,k}} ]\right]^T, m \in {{\cal M}_k}$.
	
	The first-order partial derivative of $F\left( \bf x \right)$ is given by
	\begin{equation}
		\setlength\abovedisplayskip{5pt}
		\setlength\belowdisplayskip{5pt}
		\label{partial_Fx}
		\frac{{\partial F\left( {\bf{x}} \right)}}{{\partial {x_{j,k}}}} = \frac{{\sqrt {{{\rm{e}}^{{x_{j,k}}}}} }}{{2\sum\limits_{m \in {{\cal M}_k}} {\sqrt {{{\rm{e}}^{{x_{m,k}}}}} } }} = \frac{{\sqrt {\left( {N - {t_j}} \right)p_{j,k}^d{\hat \lambda _{j,k}}} }}{{2{\theta _k}}},
	\end{equation}
	where $\rm e$ is the exponent. The second-order partial derivatives of $F\left( \bf x \right)$ are given by
	\begin{equation}
		\label{partial_Fx_1}
		\frac{{{\partial ^2}F\left( {\bf{x}} \right)}}{{\partial {{\left( {{x_{j,k}}} \right)}^2}}} = \frac{{\sqrt {{e^{{x_{j,k}}}}} \left( {\sum\limits_{m \in {{\cal M}_k}} {\sqrt {{e^{{x_{m,k}}}}} } } \right) - {{\left( {\sqrt {{e^{{x_{j,k}}}}} } \right)}^2}}}{{4{{\left( {\sum\limits_{m \in {{\cal M}_k}} {\sqrt {{e^{{x_{m,k}}}}} } } \right)}^2}}},
	\end{equation}
	and
	\begin{equation}
		\label{partial_Fx_2}
		\frac{{{\partial ^2}F\left( {\bf{x}} \right)}}{{\partial {x_{j,k}}\partial {x_{i,k}}}} = \frac{{ - \sqrt {{e^{{x_{j,k}}}}{e^{{x_{i,k}}}}} }}{{4{{\left( {\sum\limits_{m \in {{\cal M}_k}} {\sqrt {{e^{{x_{m,k}}}}} } } \right)}^2}}}.
	\end{equation}
	
	Then, we define ${{\bf{z}}_k} = {\left[ {\sqrt {{e^{{x_{1,k}}}}} ,\sqrt {{e^{{x_{m,k}}}}}, \cdot  \cdot  \cdot,\sqrt {{e^{{x_{\left| {{{\cal M}_k}} \right|,k}}}}} } \right]^T}, m \in {\cal{M}}_k$, and thus the Hessian matrix of $F\left( \bf x \right)$ can be given by 
		\begin{equation}
			\label{Hessian}
			\begin{split}
				{\bf H} &= \frac{1}{{4{{\left( {{{\bf{1}}^T}{{\bf{z}}_k}} \right)}^2}}}\left\{ {\left( {\sum\limits_{m \in {{\cal M}_k}} {\sqrt {{e^{{x_{mk}}}}} } } \right)\left[ {\begin{array}{*{20}{c}}
							{\sqrt {{e^{{x_{1,k}}}}} }&{ \cdot  \cdot  \cdot }&0\\
							0&{\sqrt {{e^{{x_{m,k}}}}} }&0\\
							0&{ \cdot  \cdot  \cdot }&{\sqrt {{e^{{x_{\left| {{{\cal M}_k}} \right|,k}}}}} }
					\end{array}} \right] - {{\bf{z}}_k}{{\left( {{{\bf{z}}_k}} \right)}^T}} \right\}\\
				& = \frac{1}{{4{{\left( {{{\bf{1}}^T}{{\bf{z}}_k}} \right)}^2}}} \underbrace {\left\{ {{{\bf{1}}^T}{{\bf{z}}_k}diag\left\{ {{{\bf{z}}_k}} \right\} - {{\bf{z}}_k}{{\left( {{{\bf{z}}_k}} \right)}^T}} \right\}}_{\bm{\Xi }},
			\end{split}
		\end{equation}

	In (\ref{Hessian}), ${\bm 1}$ is a vector of $\left[1, 1, \cdot \cdot \cdot, 1 \right]^T$, $|{\cal M}_k|$ means the cardinality of the set ${\cal M}_k$. For any given ${\bf{v}} = {\left[ {{v_1}, \cdot  \cdot  \cdot ,{v_{\left| {{{\cal M}_k}} \right|}}} \right]^T} \in {\mathbb R}^{|{\cal M}_k|}$, by using the Cauchy-Schwartz inequality, we have the inequality that is given by 
		\begin{equation}
			\label{inquality_FX}
			\begin{split}
				{{\bf{v}}^T}{\bm{\Xi} }\bf{v} &= {{\bf{1}}^T}{{\bf{z}}_k}{{\bf{v}}^T}diag\left\{ {{{\bf{z}}_k}} \right\}{\bf{v}} - {{\bf{v}}^T}{{\bf{z}}_k}{\left( {{{\bf{z}}_k}} \right)^T}{\bf{v}} \\
				& =  \left( {\sum\limits_{m \in {{\cal M}_k}} {\sqrt {{e^{{x_{m,k}}}}} } } \right)\left( {\sum\limits_{m \in {{\cal M}_k}} {{v_m}\sqrt {{e^{{x_{m,k}}}}} {v_m}} } \right) - {\left( {\sum\limits_{m \in {{\cal M}_k}} {{v_m}\sqrt {{e^{{x_{m,k}}}}} } } \right)^2} \\
				& = \left( {\sum\limits_{m \in {{\cal M}_k}} {{{\left( {\sqrt {\sqrt {{e^{{x_{m,k}}}}} } } \right)}^2}} } \right)\left( {\sum\limits_{m \in {{\cal M}_k}} {{{\left( {{v_m}\sqrt {\sqrt {{e^{{x_{m,k}}}}} } } \right)}^2}} } \right) - {\left( {\sum\limits_{m \in {{\cal M}_k}} {{v_m}\sqrt {{e^{{x_{m,k}}}}} } } \right)^2} \ge 0.
			\end{split}
		\end{equation}

	Therefore, we prove $\ln \left(\theta_k\right)$ is a convex function of $\bf x$.
	Then, by using Jensen's inequality, we have
	\begin{equation}
		\setlength\abovedisplayskip{5pt}
		\setlength\belowdisplayskip{5pt}
		\label{In_Fx}
		F\left( x \right) \ge \sum\limits_{m \in {{\cal M}_k}} {{a_{m,k}}{x_{m,k}}} + \ln\left({c_k}\right),
	\end{equation}
	where $a_{m,k}$ and $c_k$ are given in (\ref{a_k}) and (\ref{c_k}), respectively.
	
	Finally, we complete the proof by taking the exponential operation for both sides of (\ref{In_Fx}) and using $x_{m,k} = \ln \left(p_{m,k}^d\right)$.
	
\end{appendices}

\bibliographystyle{IEEEtran}
\bibliography{myref}

\end{document}